\documentclass[twocolumn,trackchanges]{aastex63}
\usepackage{amsmath,amssymb,amstext,float}

\usepackage{color}

\usepackage[all]{hypcap} 
\usepackage{aas_macros}
\usepackage{natbib}
\usepackage{fancyhdr}
\usepackage[normalem]{ulem}
\fancypagestyle{specialfooter}{%
 \fancyhf{}
 
 \fancyfoot[C]{Distribution Statement A.  Approved for public release: distribution is unlimited.}  }
\usepackage{mathtools}
\usepackage{lineno}
\shorttitle{Assessing ADAPT and AFT With In-Situ Measurements}
\shortauthors{Knizhnik et al.}
\newcommand{\beg}[1]{\begin{equation}\label{#1}}
\newcommand{\done}{\end{equation}}
\newcommand{\pd}[2]{\frac{\partial #1}{\partial #2}}

\newcommand{\at}{\bigg|}

\newcommand{\vecB}{\emph{\textbf{B}}}

\newcommand{\vecv}{\emph{\textbf{v}}}

\newcommand{\oftp}{(\theta,\phi)}

\newcommand{\curl}[1]{\nabla\times{#1}}
\newcommand{\divv}[1]{\nabla\cdot{#1}}

\newcommand{\kjk}[1]{\textcolor{black}{#1}}
\newcommand{\kjka}[1]{\textcolor{black}{#1}}
\newcommand{\kjkaa}[1]{\textcolor{black}{#1}}
\newcommand{\mjw}[1]{\textcolor{black}{#1}}
\numberwithin{equation}{section}

\begin{document}

\title{Assessing the Performance of the ADAPT and AFT Flux Transport Models Using In-Situ Measurements From Multiple Satellites}

\author[0000-0002-2544-2927]{Kalman J. Knizhnik}
\affiliation{Naval Research Laboratory, 4555 Overlook Avenue SW, Washington, DC 20375, USA}

\author[0000-0002-4433-4841]{Micah J. Weberg}
\affiliation{George Mason University, 4400 University Dr, Fairfax, VA 22030}

\author[0000-0001-8875-7478]{Elena Provornikova}
\affiliation{Johns Hopkins Applied Physics Laboratory, 11100 Johns Hopkins Road, Laurel, MD 20723, USA}

\author[0000-0001-6102-6851]{Harry P. Warren}
\affiliation{Naval Research Laboratory, 4555 Overlook Avenue SW, Washington, DC 20375, USA}

\author[0000-0002-4459-7510]{Mark G. Linton}
\affiliation{Naval Research Laboratory, 4555 Overlook Avenue SW, Washington, DC 20375, USA}

\author[0000-0002-3089-3431]{Shaheda Begum Shaik}
\affiliation{George Mason University, 4400 University Dr, Fairfax, VA 22030}

\author[0000-0002-8747-4772]{Yuan-Kuen Ko}
\affiliation{Naval Research Laboratory, 4555 Overlook Avenue SW, Washington, DC 20375, USA}

\author[0000-0002-5476-2794]{Samuel J. Schonfeld}
\affiliation{Air Force Research Laboratory, Space Vehicles Directorate, Kirtland AFB, NM 87117, USA}

\author[0000-0001-5503-0491]{Ignacio Ugarte-Urra}
\affiliation{Naval Research Laboratory, 4555 Overlook Avenue SW, Washington, DC 20375, USA}

\author[0000-0003-0621-4803]{Lisa A. Upton}
\affiliation{Space Systems Research Corporation, 1940 Duke Street Suite 200, Alexandria, Virginia 22314, USA}

\correspondingauthor{K. J. Knizhnik}
\email{kalman.knizhnik@nrl.navy.mil}

\begin{abstract}
The launches of Parker Solar Probe (Parker) and Solar Orbiter (SolO) are enabling a new era of solar wind
studies that track the solar wind from its origin at the photosphere, through the corona, to multiple
vantage points in the inner heliosphere. \kjk{A key ingredient for these models is the input photospheric magnetic field map that provides the boundary condition for the coronal portion of many heliospheric models. In this paper, we perform steady-state, data-driven magnetohydrodynamic  (MHD) simulations of the solar wind during Carrington rotation 2258 with the GAMERA model. We use the ADAPT and AFT flux transport models and quantitatively assess how well each model matches in-situ measurements from Parker, SolO, and Earth. We find that both models reproduce the magnetic field components at Parker quantitatively well. At SolO and Earth, the magnetic field is reproduced relatively well, though not as well as at Parker, and the density is reproduced extremely poorly. The velocity is \kjka{overpredicted at Parker, but not at SolO or Earth, hinting that the Wang-Sheeley-Arge (WSA) relation, fine-tuned for Earth, misses the deceleration of the solar wind near the Sun}. We conclude that AFT performs quantitatively similarly to ADAPT in all cases and that both models are comparable to a purely WSA heliospheric treatment with no MHD component. Finally, we trace field lines from SolO back to an active region outflow that was observed by Hinode/EIS, and which shows evidence of elevated charge state ratios.}
\end{abstract}



\section{Introduction}\label{sec:intro}
Understanding how the solar wind is heated and accelerated is one of the most important open questions in heliophysics. It is well established that the high-speed solar wind that permeates much of the heliosphere originates in large coronal holes \citep{Hundhausen77,Zirker77}, 
while there is also a slow-speed solar wind that appears to be partly formed by the interaction of closed field regions on the Sun with neighboring open field lines \footnote{Technically, ``closed" field lines are those that leave and return to the photospheric surface within an Alfv\'en surface (spanned by all points where $V_r=V_A$, with $V_A$ the Alfv\'en speed) since field lines extending beyond this radius (``open" field lines) cannot communicate information via Alfv\'en waves back to the surface. However, in potential field source surface models (\autoref{sec:initial_boundary_conditions} and \autoref{sec:appendix}), all field lines extending beyond the source surface are considered ``open".}, often called interchange reconnection \citep{Gosling97,Antiochos11}. Thus the properties of the solar wind are tied to its magnetic origin and tracking solar wind streams back to the solar surface is critical for advancing our understanding of plasma processes on open field lines. \par
Since the 1970s, space missions have been providing solar wind plasma and magnetic field measurements from near Earth at 1 astronomical unit (AU; 1 AU $ = 215R_\odot$, where $R_\odot = 697\;\mathrm{Mm}$) from the Sun. Satellites such as the International Monitoring Platform-8 \citep[IMP-8;][]{imp8}, International Sun-Earth Explorer-3 \citep[ISEE-3;][]{isee3}, Wind \citep{Wind}, Solar and Heliospheric Observatory \citep[SOHO;][]{soho}, and Advanced Composition Explorer \citep[ACE;][]{ace} have measured plasma properties near Earth.  The Solar Terrestrial Relations Observatory \citep[STEREO;][]{stereo} measured plasma properties off the Sun-Earth line at 1 AU. Helios \citep{helios} measured plasma properties down to 0.3 AU. Ulysses \citep{ulysses} measured plasma properties out of the ecliptic. \par 
The recent launch of Parker Solar Probe \citep[Parker; ][]{Fox16} and Solar Orbiter \citep[SolO; ][]{Muller13,Muller20} have heralded a new era for solar and heliospheric science thanks to their unique orbits and synergetic measurements from various positions in the inner heliosphere. Parker was launched on 2018 August 12 and is designed to go within 9 solar radii ($R_\odot$) of the solar surface, marking the closest approach ever of an artificial satellite to the Sun. It carries a state-of-the-art suite of instruments for both remote sensing \citep[WISPR;][]{Vourlidas16} and in-situ \citep[FIELDS;][]{Bale16}, \citep[SWEAP;][]{Kasper16}, \citep[IS$\odot$IS;][]{McComas16} measurements. Parker measurements have been used to constrain photospheric flux transport models \citep[e.g.,][]{Wallace22} and \kjka{coronal heating rates in the solar wind using simultaneous remote sensing and in-situ measurements from SolO \citep[e.g.,][]{Telloni23}}. For a recent review of Parker Solar Probe's discoveries, see \citet{Raouafi23}. \par 
SolO was launched on 2020 February 10 and will reach a perihelion of about 60 solar radii. While Parker will remain very close to the ecliptic plane, the inclination of SolO’s orbit will increase by repeated gravitational interactions with Venus. During its extended mission phase, SolO will reach latitudes of more than $30^\circ$ above the ecliptic. SolO carries suites of instruments for both remote sensing and in-situ observations such as \kjkaa{SWA \citep{Owen20} and MAG \citep{Muller13,Horbury20}}.\par 
This new era for solar and heliospheric science is further afforded by a slew of cotemporaneous solar observations from both space-based and ground-based observatories. Of particular importance are spectroscopic observations from instruments such as the EUV Imaging Spectrometer \citep[EIS; ][]{Culhane07} onboard Hinode \citep{Kosugi07} and the Interface Region Imaging Spectrograph \citep[IRIS; ][]{DePontieu14} and the Spectral Imaging of the Coronal Environment \citep[SPICE; ][]{Anderson20} onboard SolO, which are capable of detecting dynamic properties of the plasma in the solar upper atmosphere such as thermal/non-thermal motions and outflows. Imaging observations from instruments such as the Atmospheric Imaging Assembly \citep[AIA; ][]{Lemen12} onboard the Solar Dynamics Observatory (SDO), Extreme Ultraviolet Imager \citep[EUVI; ][]{Howard08} on STEREO and the Solar Orbiter Extreme Ultraviolet Imager \citep[EUI; ][]{Rochus20} provide an important context of the dynamics in the solar atmosphere. \par 
Due to the observational gap in space between the remote-sensed solar observations and the in-situ measurements in the inner heliosphere, modeling is still required to link the solar wind measured in situ back to its origin at the Sun. Most such models consist of two parts: 1) a coronal model, with the photospheric magnetic field as its inner boundary condition, spanning from the photosphere \kjka{up} to 0.1 AU, coupled with 2) an inner heliospheric model (loosely defined here as between 0.1-1 AU), where the outer boundary of the coronal model at 0.1 AU is the inner heliospheric model's inner boundary. \par 
Coronal models \kjka{actually start at the photosphere, where remote sensing instruments such as the Helioseismic and Magnetic Imager (HMI) \citep{Scherrer12} provide photospheric magnetograms, maps of the photospheric radial magnetic field, that are used as a boundary condition for the coronal model.} \kjka{Multiple magnetograms taken over an extended time period can be combined into a single ``synoptic" map. Synoptic maps are traditionally constructed by taking a strip at the central meridian from Earth over the course of a full solar rotation and stacking these up horizontally to create an approximate map of the full sun over that time period \citep{Harvey80,Ulrich06}}. 
\kjka{Since the photospheric magnetic field is continuously evolving, techniques to dynamically evolve flux have been developed in order to transport flux self-consistently around the solar surface. This permits the creation of a ``synchronic map”, which shows the full Sun at a single instant in time, where the Earth side view of the Sun is a Carrington projected assimilation of a full disk magnetogram of that instant, and the remaining unobserved area is populated with transported magnetic flux from a synchronic map of a previous instant. These flux transport models typically differ in how new data is assimilated into the map, and what physical processes are modeled \citep{Arge10,Upton14}. To a large extent, as will be described below, the output of coronal and heliospheric modeling is only as good as these input flux transport maps. It is therefore crucial to determine the most accurate boundary condition for the coronal models.} 
\kjk{Each} coronal model, in turn, can have various degrees of complexity in their treatment of coronal physics, ranging from potential field source surface \citep[PFSS; ][]{Altschuler69, Schatten69} models to magnetohydrodynamic (MHD) models. PFSS models are typically combined with a Schatten current sheet (SCS) model to connect the source surface \citep[usually defined as a sphere concentric on the Sun at a distance around 2.5$R_\odot$ from Sun center;][]{Altschuler69} to the inner heliosphere and beyond. The Wang-Sheeley-Arge model \citep[WSA; ][]{Wang90,Arge00} is used to relate the magnetic field structure inside the source surface to the plasma properties at and beyond 0.1 AU. Using WSA\footnote{In this paper, the WSA model will henceforth be assumed to include the SCS in the coronal portion of the model. Thus, writing `WSA' implicitly assumes the presence of the SCS from $R_{ss}$ outward.}, therefore, enables models to bypass running a full MHD simulation out to 0.1 AU. Meanwhile, coronal MHD models such as MAS/CORHEL \citep{Riley12,Lionello14}, AWSoM \citep{vanD14}, \kjka{COCONUT \citep{Perri22}} and MULTI-VP \citep{Pinto17} solve the MHD equations inside 0.1 AU and can be fed directly into inner heliospheric models.\par 
Inner heliospheric models can also have various degrees of complexity in their treatment of inner heliospheric physics, ranging from ballistic propagation to 
MHD simulations. In ballistic models \citep{Nolte73,Badman20,Badman21,Badman22,Badman23}, the magnetic field outside \kjka{the coronal model} is taken to follow a Parker spiral so that magnetic field lines can be traced from anywhere in the heliosphere ballistically back to 0.1 AU, and then to the photosphere. Meanwhile, MHD models such as the inner heliospheric versions of MAS and AWSoM, as well as EUHFORIA \citep{Pomoell18}, or ENLIL \citep{Odstrcil03} predict the inner heliospheric environment of the solar wind and the propagation of CMEs within it by taking the output from coronal models as input and are also well-suited for tracing the solar wind/open field from any location back to its footpoint at the solar surface. \par 
Previous work has tested various models against each other using various statistical techniques \citep{MacNeice11,Riley13,Samara21,Jivani22}, and has produced impressive comparisons between in-situ Parker data and single-fluid MHD models \citep{Kim20,Pogorelov20,Riley21,vanD22}, two fluid MHD models \citep{Chhiber21}, as well as ballistic models \citep{Nolte73,Badman20,Badman21,Badman22,Badman23}. MHD models have also been used successfully to predict CME signatures at 1 AU \citep{Wu97,Wu07,Wu11,Wu16a,Wu16b,Wu20}.\par
One important challenge in any model that relates remote sensing observations with in-situ predictions or measurements of solar wind properties is accurately connecting the in-situ data with its source location on the Sun. The most direct way to do this is by tracing magnetic field lines in order to map plasma from the inner heliosphere to the photosphere. In practice, this mapping depends on the combination of models used to extrapolate the photospheric magnetic field into the inner heliosphere. A recent study \citep{Badman23} compared four different approaches to obtaining the source locations of plasma measured by Parker during \kjka{its $10^{th}$ perihelion pass}. The four approaches were 1) a pure WSA model out to $21.5 R_\odot$ with a ballistic heliospheric model beyond that, 2) a PFSS model to the source surface and a ballistic heliospheric model beyond that, 3) a MHD/ballistic model inside/outside $30 R_\odot$ and 4) a WSA model to $10 R_\odot$ and an MHD model outside $10R_\odot$. The authors find that the models result in approximately the same source regions when solar activity is low and there are plenty of equatorial coronal holes to warp the heliospheric current sheet away from Parker so that small differences between the models do not encounter any magnetic separatrices.\par  
In this paper, we present a new coupled coronal and inner heliospheric model suite, namely WSA+Grid Agnostic MHD for Extended Research Applications \citep[GAMERA; ][]{Gamera,Sorathia20,Mostafavi22}, and model the evolution of the heliosphere during Carrington Rotation (CR) 2258. One major \kjk{difference between our model and} previous iterations of the WSA+MHD combination of models \citep[e.g.,][with ENLIL]{Odstrcil03}, \citet{Wu16a,Wu16b,Wu20} with the H3DMHD code and \citet{Kim20} with the MS-FLUKSS code as the MHD component of the coupled model is our use of the Advective Flux Transport (AFT) model \citep{Upton14,UgarteUrra15,Upton18}, in addition to the commonly used Air Force Data Assimilative Photospheric Flux Transport (ADAPT) model \citep{Arge10,Arge13,Hickmann15} to seed the magnetic field and velocity state of the low corona and inner heliosphere. 
\kjk{In this paper, we use both ADAPT and AFT to obtain predictions at various satellites in the inner heliosphere. We quantitatively compare MHD and WSA heliospheric models whose coronal portion is determined from the ADAPT and AFT models by studying the in-situ measurements at Parker, SolO, and Earth. Finally,} we make use of the composition instrumentation on SolO \kjkaa{\citep[SWA-HIS;][]{Livi23}} to map and validate SolO's magnetic connectivity to a source region on the Sun that was observed in a Hinode/EIS raster. Validation of these connectivity models is critical for understanding solar wind acceleration and heating.\par
This paper is organized as follows. In \autoref{sec:model} we describe our models connecting the photosphere to the source surface, 0.1 AU, and then 1 AU. In \autoref{sec:Results}, we compare the model results with the in-situ measurements by Parker, SolO, and spacecraft at Earth and trace the open magnetic field lines from these spacecraft locations to the source regions of the solar wind. In \autoref{sec:conclusions}, we discuss some of the implications and interpretations of our results.

\section{Numerical Model}\label{sec:model}
\subsection{The GAMERA Model}\label{subsec:model}
GAMERA is an MHD solver with numerical algorithms inherited from the Lyon-Fedder-Mobary \citep[LFM; ][]{Lyon04} MHD code which features high-order spatial reconstruction in arbitrary non-orthogonal curvilinear coordinates and a constrained transport method fulfilling the $\nabla \cdot \mathbf{B}=0$ condition to machine precision. Following the application of the LFM code to simulations of the inner heliosphere \citep{merkin2016coupling, Merkin16, Merkin_etal_2011}, it was rewritten as GAMERA for magnetospheric applications. GAMERA, in turn, has recently been used to model the solar wind in the steady state inner heliosphere to study the generation of the Kelvin-Helmholtz instability \citep{Mostafavi22}. Compared to LFM, GAMERA includes upgrades in the numerical scheme, improvements in code implementation, and hybrid parallelism using MPI and OMP. GAMERA was designed to be portable and optimized for modern supercomputer architectures while maintaining the algorithmic features of LFM.\par
\par 
GAMERA solves the semi-conservative ideal MHD equations having the form:
\beg{massconservation}
\pd{\rho}{t} = -\divv{(\rho \vecv)},
\done 
\beg{momentumconservation}
\pd{\rho\vecv}{t} = -\divv{\Big(\rho\vecv\vecv+\textbf{I}P\Big)} - \divv{\Big(\textbf{I}\frac{B^2}{2}-\vecB\vecB\Big)},
\done 
\beg{energyconservation}
\pd{W}{t} = -\divv{\Big(\vecv(W+P)\Big)} - \vecv\cdot\divv{\Big(\textbf{I}\frac{B^2}{2}-\vecB\vecB\Big)},
\done 
and
\beg{induction}
\pd{\vecB}{t} = \curl{\Big(\vecv\times\vecB\Big)}.
\done 
The variables have their usual meanings: $t$ is time, $\rho$ and $P$ are the plasma density and thermal pressure, $\vecv$ and $\vecB$ are the velocity and magnetic field vectors, respectively, and $\textbf{I}$ is the identity matrix. The energy $W$ is defined as
\beg{defW}
W = \frac{1}{2}\rho v^2 + \frac{P}{\gamma-1},
\done 
for the ratio of specific heats, $\gamma=5/3$. \kjka{The model ignores heat flux and wave pressure in the energy equation}. A complete description of the numerical method of solving these equations can be found in \citet{Gamera}.
\subsection{Coordinate Systems}
The native output of the GAMERA code is an irregularly gridded Cartesian data set. After each simulation, we perform a coordinate transformation to the spherical expression of the HelioCentric Inertial (HCI) coordinate system. In the HCI system \citep{Franz02}, the XY-plane (0$^{\circ}$ latitude) is aligned with the solar equatorial plane observed at 2000-01-01 12:00 UTC, the X-axis (0$^{\circ}$ longitude) points towards the solar ascending node on the ecliptic of J2000.0, and the Z-axis (+90$^{\circ}$ latitude) points along the solar rotation axis. The RTN spherical system used below to describe the simulation results \citep{Franz02} is a local coordinate frame where $R$ is the Sun-to-spacecraft direction, $T$ is the toroidal direction obtained when the Sun's spin rotation axis is crossed with $R$, and $N$ completes the right-handed system. Here we use spherical coordinates r, theta, phi (radius, colatitude, longitude), where solar north is $\theta=0$ and the equator is \kjka{$\theta=\pi/2$}, where appropriate. To be consistent with in-situ notation, we use R, T, N is for in-situ and modeled in-situ data. In these coordinate systems, R is interchangeable with r.  In this paper, we will use spherical coordinate subscripts $r,\theta,\phi$ to denote quantities at radii less than or equal to the radius of the GAMERA inner boundary, and $R, T, N$ to subscript variables modeled by GAMERA.


\subsection{Initial and Boundary Conditions}\label{sec:initial_boundary_conditions}
\begin{figure*}[ht]
\includegraphics[trim={0cm 0cm 0cm 0cm},clip,scale=0.6]{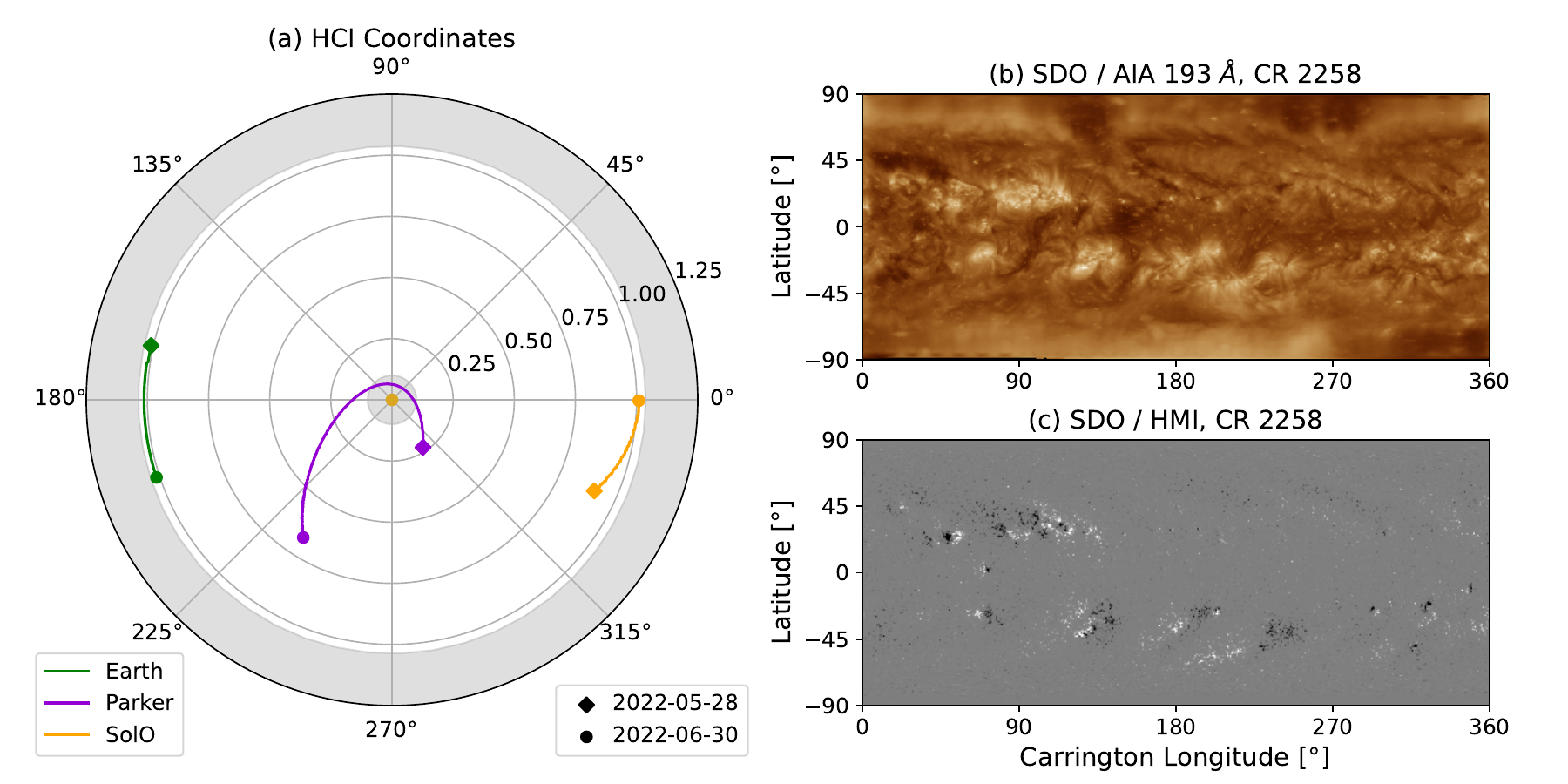}
\caption{Overview of the modeled time period. Panel (a) shows the trajectories of Earth, Parker, and Solar Orbiter during the time frame, and (b) and (c) show, respectively, synoptic maps of AIA 193 \AA $\;$ and HMI observations from SDO. In (a), white/gray indicates regions inside/outside our computational domain (see \autoref{sec:model}).}\label{fig:overview}
\end{figure*}

For this study, we examine \kjk{Carrington rotation 2258 (CR2258; 2022 May 28 -- 2022 Jun 24), which included the period of Parker Encounter 12
from 2022 May 27 to 2022 June 7}. \autoref{fig:overview} shows the trajectories of Earth, Parker, and SolO during this time, along with synoptic maps of AIA 193 \AA $\;$ and HMI observations from SDO. \kjk{The HMI map was downloaded from the Joint Science Operations Center (JSOC; http://jsoc.stanford.edu/) while the AIA map was generated by using 55 full-sun maps ($\sim$12 hr cadence). For each AIA image, we applied a Gaussian filter with a full width at half maximum of 7.8 degrees centered on the central meridian. Then, we reprojected each filtered image into the heliographic (Carrington) coordinate frame and summed the intensities.} 

In this study, we employ synchronic maps from two photospheric flux transport models to populate the input magnetic field map for CR2258: the ADAPT and AFT models. ADAPT is based on the \citet{Worden00} flux transport method and uses data assimilation techniques to incorporate data from new magnetograms. Measurements near the central meridian (limb) are given the most (least) weight since they are the most (least) reliable. \kjk{AFT also uses data assimilation to incorporate magnetic data, but while ADAPT typically only assimilates data within $60^{\circ}$ from disc center, AFT assimilates data from the input magnetogram to within 3 pixels of the limb.} ADAPT also includes the effect of random flux emergence. Both models include differential rotation and meridional flow. \kjk{AFT uses the observed meridional flow and differential rotation measured by \citet{Hathaway_etal2022}, whereas the ADAPT meridional flow profile takes the form given by \citet{Wang94} and is tuned so that the model reproduces the polar field evolution. Both models incorporate motions representative of the convective flows in some way. ADAPT uses a stochastic
diffusion method to mimic the supergranular motions \citep{Simon95}. In contrast, AFT models the surface flows using vector spherical harmonics to create a convective velocity field that reproduces the  characteristics (e.g., size, lifetimes, velocities) of the convective flows observed on the Sun \citep{Hathaway10,UgarteUrra15}. }

 \par 
Both models are steady-state in our simulations, meaning that the input map does not get updated during the simulated time period. Improving this assumption will be the subject of future work. \par 
To model the solar wind in the inner heliosphere, GAMERA requires inner boundary conditions at some height, set here to be 0.1 AU or 21.5 $R_\odot$. In previous studies, these have been provided by the empirical Wang-Sheeley-Arge \citep[WSA; ][]{Wang90,Arge00} coronal model based on ADAPT input magnetic fields \citep{Mostafavi22}, but in principle can be provided by, for example, WSA based on AFT input, or even by a full MHD model of the low corona \citep[e.g.; ][]{Lionello14,vanD14}. Our implementation of WSA in this paper can use either ADAPT or AFT as the photospheric boundary condition. \par 
The WSA model relies on a coronal magnetic field, modeled by a PFSS extrapolation solution of a Carrington map of the photospheric magnetic field into the corona to a height of typically a few $R_\odot$. The PFSS solution \citep{Altschuler69} assumes that the low corona is current free between the photosphere (where the radial magnetic field is measured directly) and the `source surface' at $R_{ss} = 2.5\; R_\odot$ (where the field is assumed to be purely radial). The field is then extrapolated from the source surface to 0.1 AU with a SCS model. The SCS solution \citep{Schatten71} assumes the corona outside $R_{ss}$ is current-free between the source surface and infinity, except for a heliospheric current sheet (HCS) separating the positive and negative polarities of the radial magnetic fields. The description of our implementation of the PFSS and SCS solution can be found in \autoref{app:intro}-\autoref{app:scs}.\par 
We obtain the distribution of the radial velocity in the inner heliosphere via an empirical formula from WSA that uses information on the expansion of the magnetic field flux tubes between the photosphere and the source surface, and their proximity to the coronal hole boundary on the photosphere \citep{Arge00}. Together, the PFSS, SCS and the WSA velocity relationship create a model of the magnetic field and velocity from the photosphere into the inner heliosphere, and the inner boundary condition of GAMERA can be extracted from a slice from this solution at some radius, chosen here to be 0.1 AU. \kjka{The remaining MHD variables at the inner boundary condition are obtained as will be described below, in \autoref{densvel}-\autoref{pressurebalance}}. \par 
Before initialization of the GAMERA simulation, the coronal model output at the inner boundary of GAMERA at arbitrary resolution is interpolated to the GAMERA inner boundary grid, also with arbitrary resolution. To account for the solar rotation, at each time step during the simulation, the boundary conditions (radial and toroidal components of the magnetic field, radial component of the velocity, the density, and the temperature) are interpolated to the inertial GAMERA inner boundary grid according to an angular velocity determined by a latitudinally constant (in the simulation) sidereal rotation period of $T_s=25.38$ days \citep{Merkin_etal_2011},
corresponding to a solar latitude of about $26^\circ$ \citep[sidereal rotation is the time it takes for the Sun to make a single revolution around its axis of rotation relative to the background stars;][]{Rosa95}. In GAMERA, rigid rotation of the corona is assumed out to $21.5 R_\odot$.  \par
\begin{figure*}[ht]
\includegraphics[trim={0cm 0cm 0cm 0cm},clip,scale=0.7]{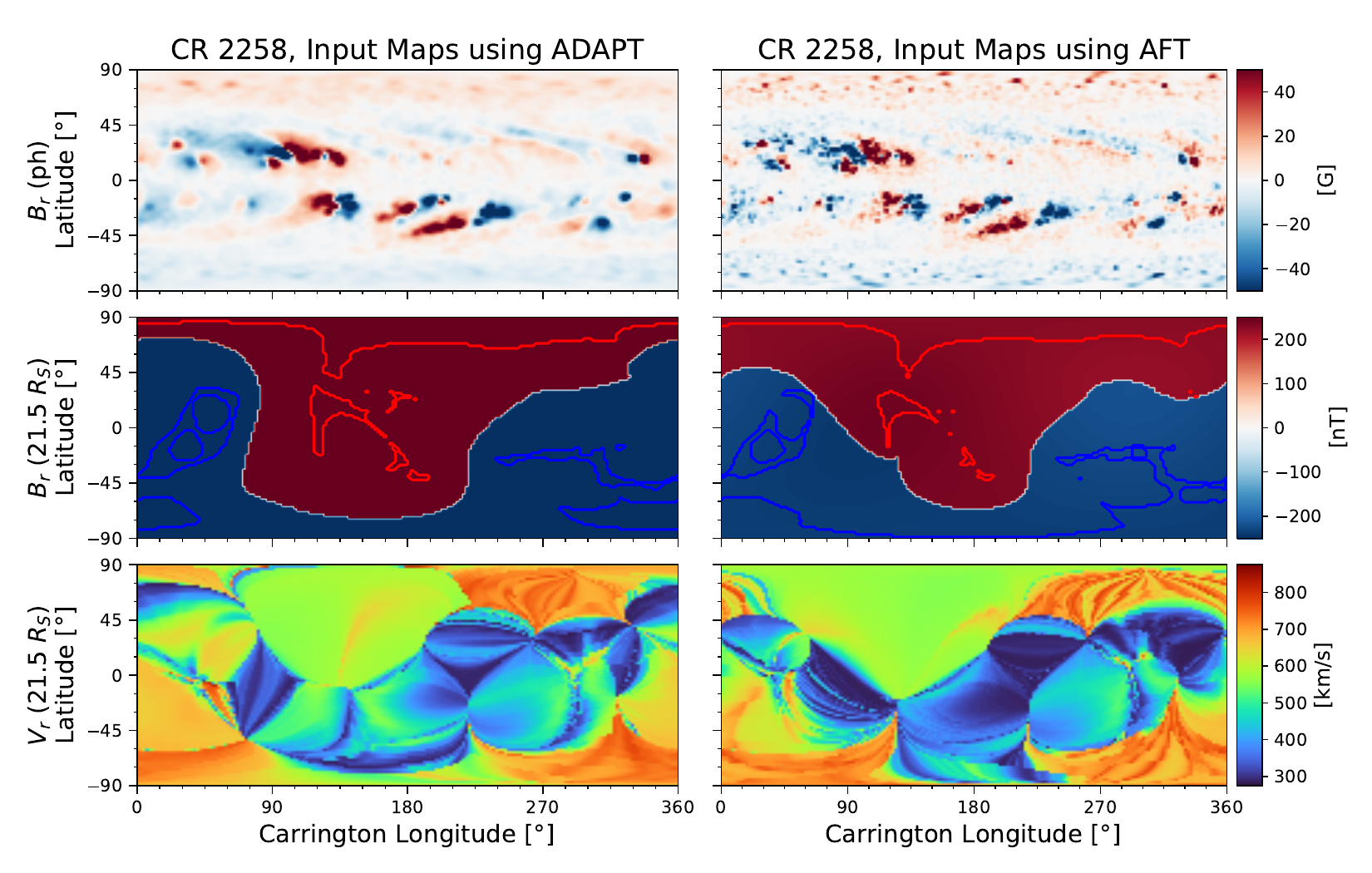}
\caption{Synchronic maps from the ADAPT (left) and AFT (right) models. The top panels show the photospheric radial magnetic field. The middle and bottom panels show the radial magnetic field and radial velocity, respectively, at the inner GAMERA boundary. These are obtained by extrapolation via a PFSS solution to the source surface followed by a SCS model out to the inner boundary. In the middle panels, red/blue lines indicate contours of open flux at the photosphere overplotted on color scale $B_r$ at $21.5 R_\odot$.}\label{fig:inputmaps}
\end{figure*}

 
\autoref{fig:inputmaps} shows the photospheric input field using the two models, as well as $B_r$ and $v_r$ at the inner GAMERA boundary $R_{in}$. The maps are synchronic, meaning they represent a single instant in time, with far-side information updated to that instant in time according to the ADAPT and AFT models. For this study, we use an ADAPT map produced at 2022-06-10 20:00 UT from the Global Oscillation Network Group (GONG) line-of-sight (LOS) data \citep{Harvey96}. The AFT map from 2022-06-10 at 16:00 UT was made using SDO/HMI LOS data. In both cases, the LOS component of the magnetic field was divided by the cosine of the angle from the disk center to approximate the radial magnetic field \citep{Upton14}. These dates and times are slightly before the midpoint of CR2258 on 2022-06-10 21:21 UT, and were chosen because GAMERA assumes that the Earth is at $180^\circ$ Carrington longitude at $t=0$. \kjk{The ADAPT model typically scales input magnetograms by an overall constant whose value depends on each magnetograph \citep{Barnes23}, which can vary over time depending on instrumental upgrades. The scaling is applied to match the photospheric flux to that measured by Kitt Peak Vacuum Telescope (C. Henney, personal communication). This scaling is, therefore, different for each magnetograph, since the strength of the magnetic field that these measure can be significantly different \citep{Virtanen17}. AFT does not natively perform this scaling, so the photospheric AFT map used in this study, shown in \autoref{fig:inputmaps}, has been multiplied by a factor of 1.58 to obtain the same net unsigned photospheric flux as in the ADAPT map.}


The maps at $R_{in}$ (middle panels) are created by performing a PFSS extrapolation of CR2258 using the ADAPT/AFT maps to $R_{ss}$, and then calculating the SCS model (\autoref{app:scs}) and extracting the magnetic field components at $R_{in}$. We initialize our $v_r$ inner boundary conditions at $R_{in}$ from the WSA velocity equation applied to the PFSS extrapolation of CR2258. In principle, $R_{in}$ can be reduced somewhat to capture regions closer to the Sun, but the inner boundary conditions of the inner heliospheric model, \kjka{which assume no radially inward characteristics} imply that flows slower than the fast magnetosonic speed, such as those occurring closer to the Sun, \kjka{will produce numerical, non-physical artifacts at the boundary}, preventing $R_{in}$ from being placed arbitrarily close to the Sun. At the inner boundary, the initial radial velocity in WSA is given by \citep[e.g., ][]{Arge00,Wallace20,Wallace21}:
\beg{WSA}
v_r = V_{sw} + \frac{V_{fw}}{(1+f_s)^{2/9}}\Bigg(1-\frac{4}{5}\exp\Big({-(d/2)^2}\Big)\Bigg)^3
\done 
where $V_{sw}$ = 285 $\mathrm{km\;s^{-1}}$ and $V_{fw}$ = 625 $\mathrm{km\; s^{-1}}$, and $f_s$ and $d$ are, respectively, the coronal magnetic flux tube expansion factor and the shortest angular distance (in degrees, calculated in the heliocentric coordinate system, described below) to the nearest coronal hole boundary at the photosphere. The expansion factor is defined as the scaled ratio of the total magnetic field magnitude at the photosphere and source surface and is given by the equation \citep{Wang90}:
\beg{fs}
f_s = \left(\frac{R_{\odot}}{R_{ss}}\right)^2 \left(\frac{B_{ph}}{B_{ss}}\right).
\done
Since magnetic flux is conserved, a self-similarly expanding flux tube has $f_s=1$, while an over (under) expanding flux tube has $f_s$ greater than (less than) one. We calculate $f_s$ and $d$ at the photosphere, then obtain $v_r$ from \autoref{WSA} by tracing field lines from $R_{in}$ to $R_{ss}$ and then transfer $v_r$ out to $R_{in}$ by assuming these are constant on field lines. The remaining velocity components at $R_{in}$ are given by
\beg{vtheta_init}
v_{\theta}(R_{in},\theta,\phi)=0,
\done 
\kjk{and
\beg{vphi_init}
v_{\phi}(R_{in},\theta,\phi) = 0, 
\done 
in the inertial frame used by GAMERA}.
Finally, the remaining magnetic field components at $R_{in}$ are given by
\beg{Btheta_init}
B_\theta(R_{in},\theta,\phi)=0,
\done 
\beg{Bphi_init}
B_\phi(R_{in},\theta,\phi) = \frac{ \Omega R_{in} \sin\theta} {v_r(R_{in},\theta,\phi)}B_r(R_{in},\theta,\phi),
\done 
where
\beg{Omega}
\Omega=2\pi/T_{s}.
\done 
The initial density is specified using an empirical relationship with the radial velocity, obtained by fitting  the Helios data \citep{McGregor11,Merkin16} as follows:
\beg{densvel}
n = 112.64 + 9.49\times 10^7 / v_r^2,
\done 
for velocity measured in $\mathrm{km\; s^{-1}}$ and density ($n=\rho/m_p$ for $m_p$ the proton mass) in $\mathrm{cm^{-3}}$. 
The initial solar wind temperature is derived assuming a balance of the total pressure (a sum of the thermal pressure and the magnetic pressure) at $R_{in}$ according to:
\beg{pressurebalance}
nk_BT +\frac{B_r^2}{8\pi} = n_{HCS}k_BT_{slow},
\done 
where $k_B$ is Boltzmann's constant. The right-hand side of the equation is a constant corresponding to the pressure in the HCS, where the magnetic pressure component is negligible. \kjka{Following \citet{Mostafavi22}, $n_{HCS}$ is the maximum number density in the HCS at $R_{in}$, and $T_{slow}=3.6\times10^5\;\mathrm{K}$ is \kjkaa{a representative temperature of the slow wind}, in line with Helios measurements \citep{SD16}.} \par 
In this simulation, the grid setup is similar to the previous GAMERA- and LFM-studies referred to in \autoref{subsec:model}. A uniform spherical grid is used in the simulation domain extending from $R_{in} \leq r \leq 220\;R_\odot$ in the radial direction, $18^\circ \leq \theta \leq 162^\circ$ in the colatitudinal direction ($\theta=0^\circ$ corresponds to the solar north), and $0 \leq \phi \leq 360^\circ$ in the longitudinal direction. The grid resolution used in this study is $N_r \times N_{\theta} \times N_{\phi} = 256 \times 128 \times 256$ leading to the cell size $\Delta r \times \Delta \theta \times \Delta \phi = 0.78 R_\odot \times 1.1^{\circ} \times 1.4^{\circ}$. \par 
The plasma variables in the simulation volume are obtained by assuming that for the initial state of the simulation volume $B_r(r)$ falls off from the inner boundary as $r^{-2}$, and the radial velocity, density, and temperature are all constant with radius, with values determined at the inner boundary of GAMERA. This causes the solar wind to blow out and equilibrate the initially uniform heliosphere over the course of about $200$ hours. Following this `negative time', at simulation time $t=0$, the $\phi$ placement of the Earth relative to the Sun is such that it will be oriented at $\phi = \pi$ at the central time of the Carrington map.


\section{Results}\label{sec:Results}
\begin{figure*}[ht]
\includegraphics[trim={0cm 0cm 0cm 3cm},clip,scale=0.2]{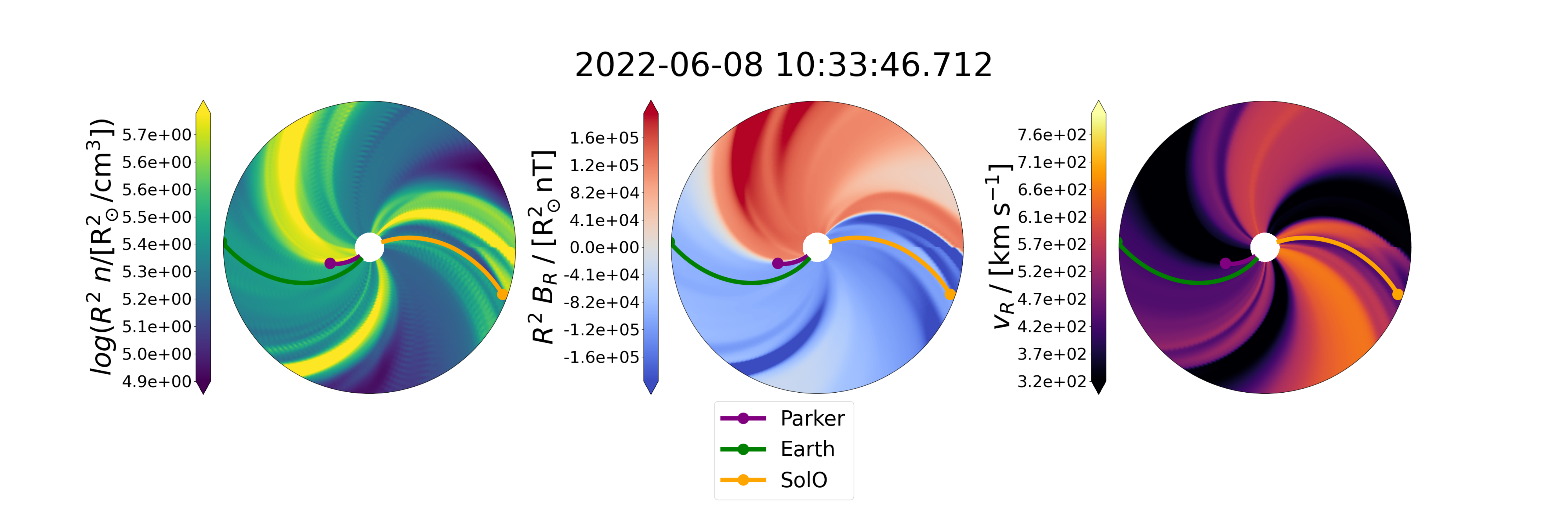}
\includegraphics[trim={0cm 7.1cm 0cm 3cm},clip,scale=0.2]{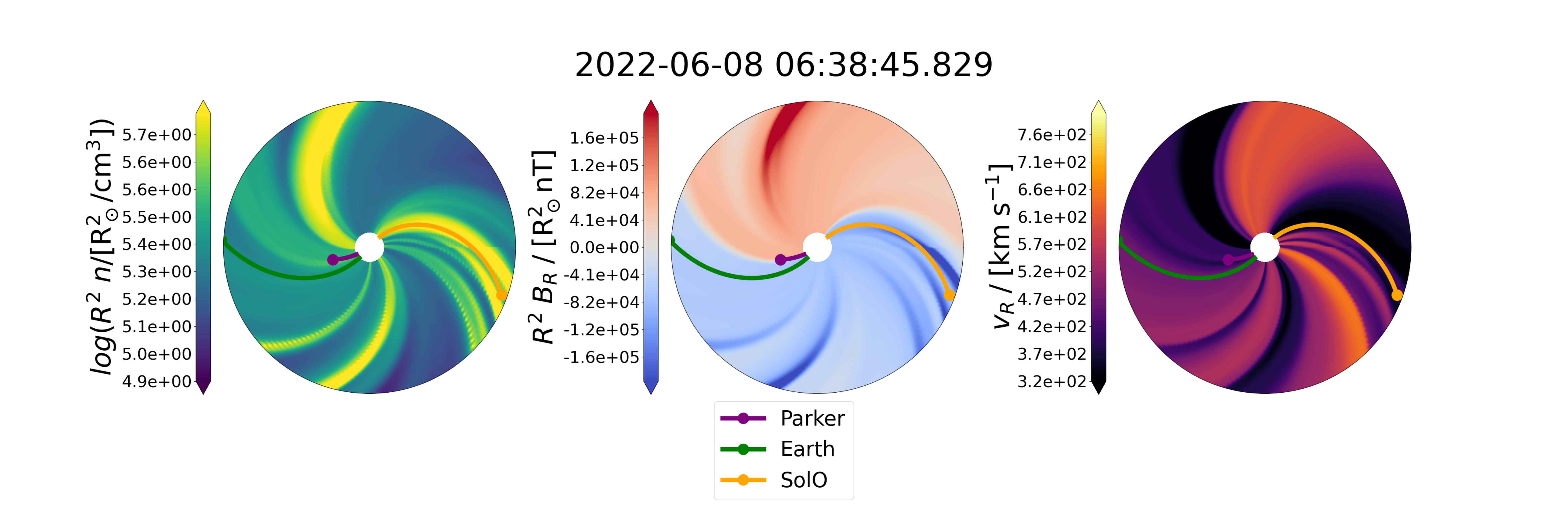}
\caption{Equatorial cuts of density (left), $B_R$ (center), and $v_R$ (right) from the GAMERA simulations using the ADAPT (top) and AFT (bottom) models. Field lines connecting the spacecraft and Earth to the inner boundary are projected onto the equatorial plane.}\label{fig:helioplots}
\end{figure*}

We plot snapshots of the heliospheric variables in \autoref{fig:helioplots}, where we show log$(R^2n)$, $R^2B_R$ and $v_R$ in the equatorial slice of GAMERA simulation output for both the ADAPT and AFT photospheric model inputs at a time near the closest approach of Parker to the Sun. \kjk{Hereafter, these will be referred to as GAMERA/ADAPT and GAMERA/AFT to differentiate them from the purely WSA treatment based on the ADAPT and AFT maps, which will be called WSA/ADAPT, WSA/AFT.} In the plot, $R$ is the radial coordinate between the Sun and each satellite. The perihelion itself ($13.3 R_\odot$ for Encounter 12) was not captured by our GAMERA runs because it was inside $R_{in}$, though this is not crucial for testing the fidelity of our model. We choose this scaling for our plots because radial magnetic and mass flux are expected to be constant, so $B_R$ and $n$ will scale as $R^{-2}$, assuming $v_R$ is approximately constant with $R$.\par 

We obtained the position of various spacecraft and Earth at each time step using the SpiceyPy \citep{Annex20} and Astropy \citep{Astropy18} packages. The state variables obtained by solving \autoref{massconservation}-\autoref{induction} were then linearly interpolated from the $8$ nearest grid points to the local position of each satellite while field lines were traced from the location of each satellite using a fourth-order Runge-Kutta solver, solving the field line equations in spherical coordinates:
\beg{fieldlineeqns}
\begin{aligned}
\frac{dr}{ds} &= \frac{B_r}{|\vecB|} \\
\frac{d\theta}{ds} &= \frac{1}{r}\frac{B_\theta}{|\vecB|}\\
\frac{d\phi}{ds} &= \frac{1}{r\sin\theta}\frac{B_\phi}{|\vecB|}.
\end{aligned}
\done 
We used fixed stepsizes of $ds=10^{-3} R_\odot$ to obtain the positions $(r_i,\theta_i,\phi_i)$ of successive points $i$ along the field line originating at each satellite at each time step. The endpoint location of the field line at each time step on the inner boundary was identified and mapped back down to its source region via the corresponding coronal magnetic field model.

\subsection{In-situ Measurements}
The in-situ solar wind data was obtained from the science-grade observations available through NASA's CDAWeb data repository. \mjw{For most of Parker's orbit, we use level-3 plasma moments from the SPC instrument \citep{Case20}, since it has the best overall cadence and quality. Unfortunately, however, SPC has saturation and field-of-view issues near perihelion (c.f. the data release notes). Therefore we fill in the missing data inside 22 $R_\odot$ using level-3 plasma moments from the SPAN-I instrument \citep{Livi22}, which has the opposite problem (i.e. good data near perihelion, but poor field-of-view and statistics further out). Throughout the entire Parker orbit, we use 1-minute magnetic field data from the MAG instrument in the FIELDS suite \citep{Bale16}. At SolO, we use bulk plasma data from SWA-PAS \citep{Louarn20} and 1-minute magnetic field observations from the MAG instrument. At Earth, we use both plasma and magnetic field data from the hourly OMNI database \citep{Papitashvili20} which aggregates data from multiple near-Earth spacecraft and time-shifts the data to the Earth's bow shock. For this time period, the OMNI data is based on plasma data from Wind / SWE \citep{Ogilvie95} and magnetic field measurements from Wind / MFI \citep{Lepping95}}. The data at all locations were resampled to a cadence of 2.5 hours \kjk{by applying a boxcar average to} only the data with ``good" quality flags.

\begin{figure*}[ht]
\includegraphics[trim={0cm 1cm 0cm 0cm},clip,scale=0.8]{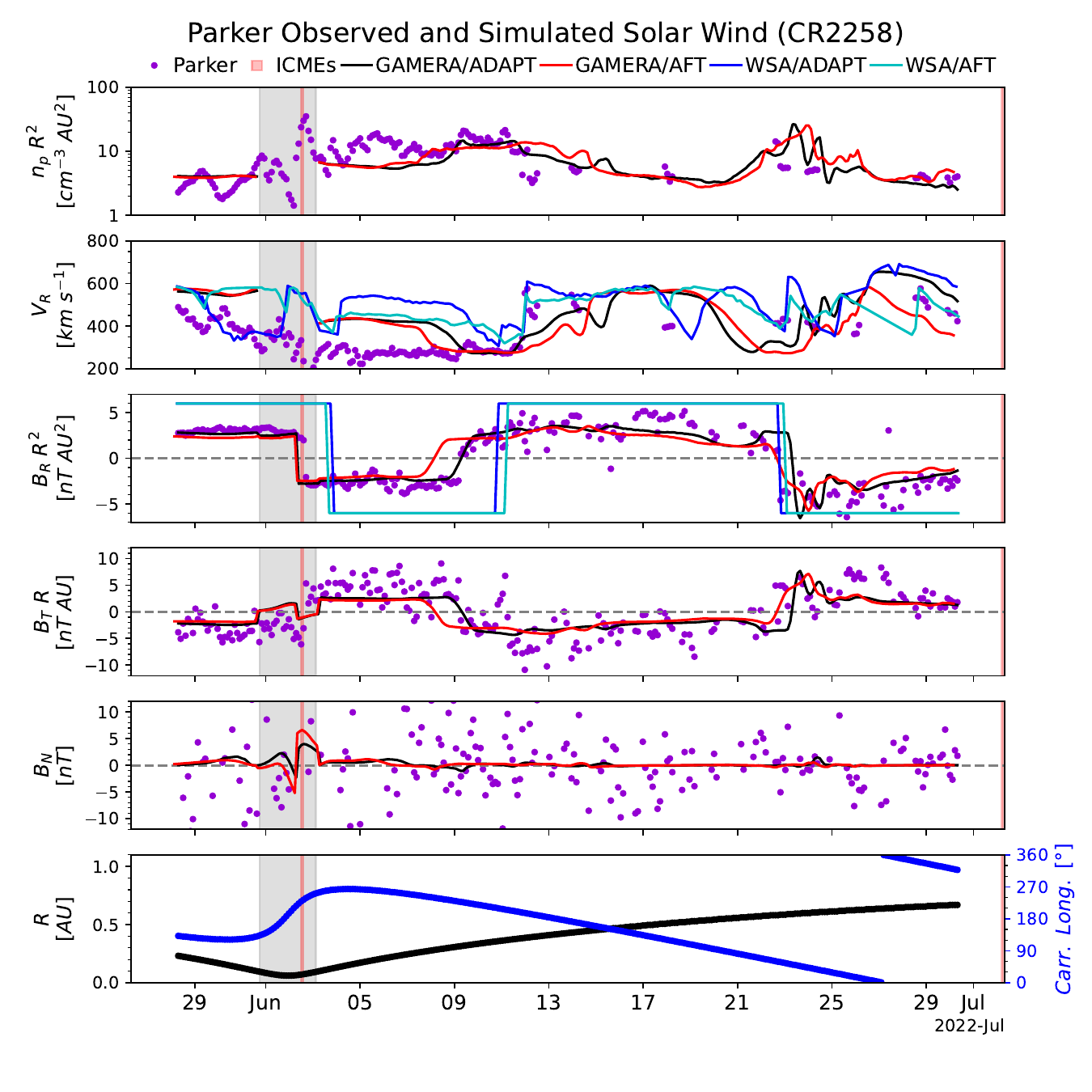}
\caption{Comparison of the measured in-situ data at Parker with the in-situ data from both GAMERA models. Density, $v_R$, $B_R R^2$, $B_T R$, $B_N$, and distance from the Sun's center ($R$) and Carrington longitude are plotted from top to bottom. ADAPT/AFT data is shown in black/red. The measured in-situ data was averaged onto the simulated $2.5\;\mathrm{hr}$ cadence. The shaded region near 01 Jun corresponds to the time interval when Parker was inside $R_{in}$, so the magnetic field values there are taken directly from the SCS layer of the coronal model. The radial velocity and magnetic field polarity from WSA/ADAPT and WSA/AFT are also shown for each case. Shaded red regions are time intervals with ICMEs measured by Parker.}\label{fig:comparepsp}
\end{figure*}

\begin{figure*}[ht]
\includegraphics[trim={0cm 1cm 0cm 0cm},clip,scale=0.8]{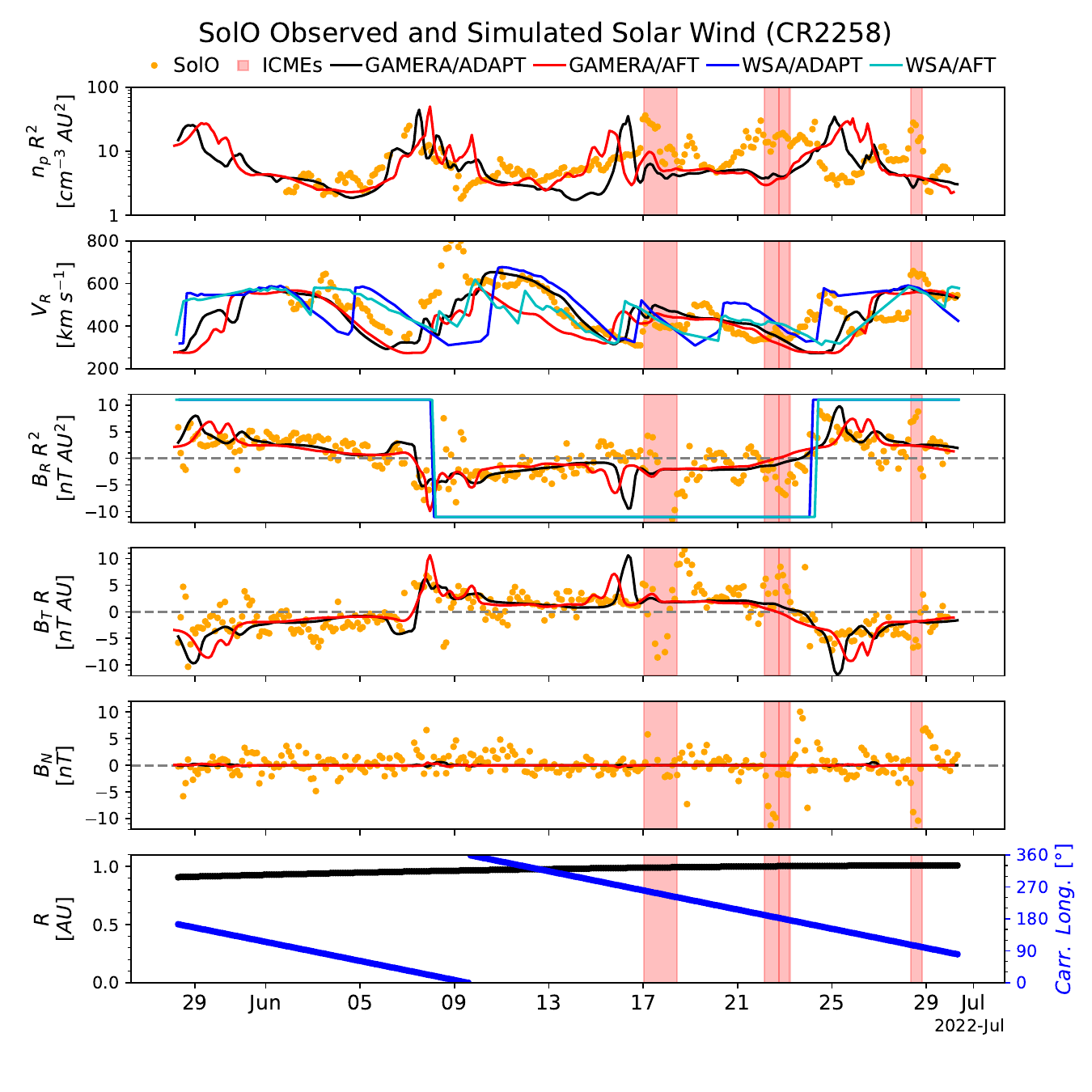}
\caption{Same as \autoref{fig:comparepsp} but for SolO. }\label{fig:compareSolO}
\end{figure*}

\begin{figure*}[ht]
\includegraphics[trim={0cm 1cm 0cm 0cm},clip,scale=0.8]{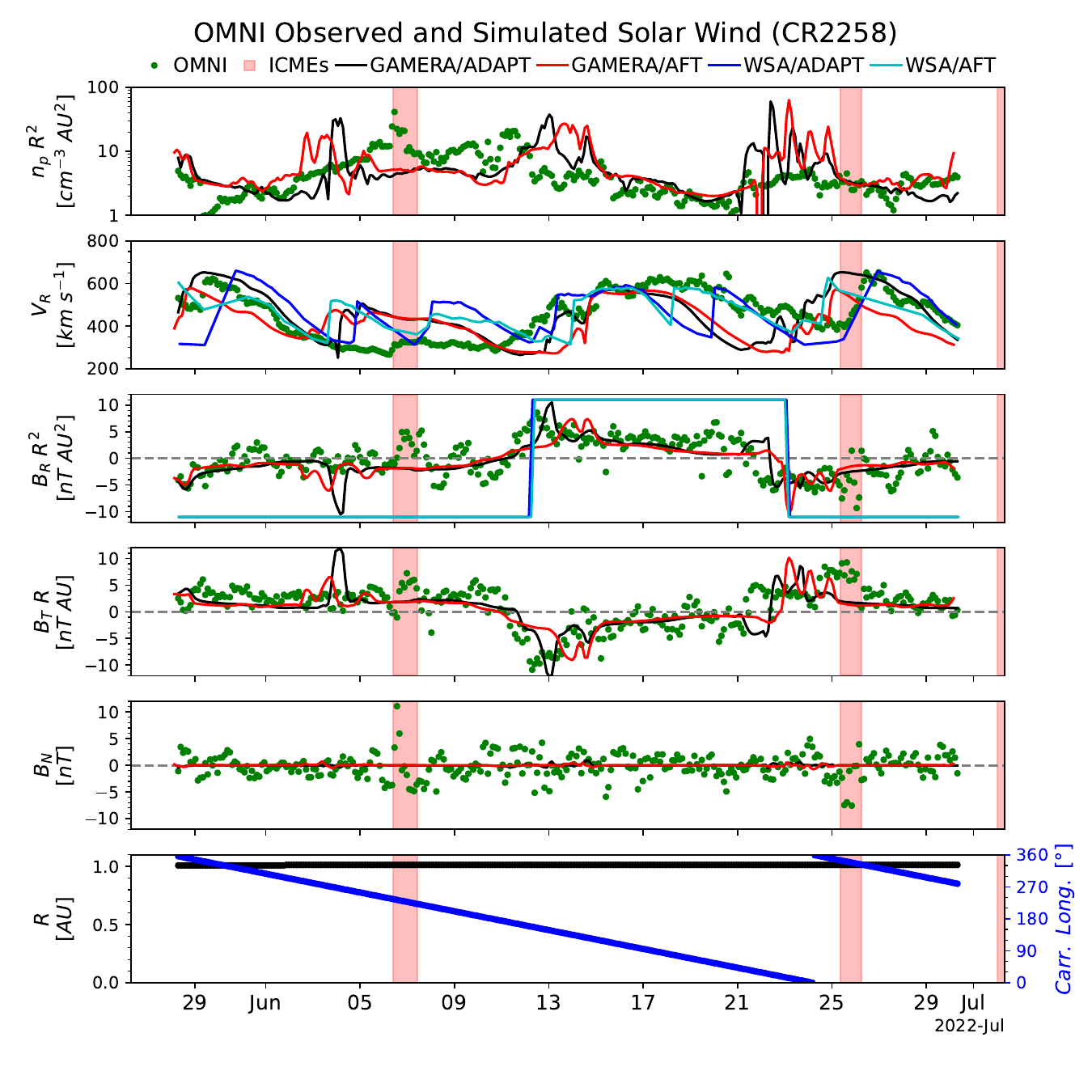}
\caption{Same as \autoref{fig:compareSolO} but for Earth.}\label{fig:compareOMNI}
\end{figure*}

\kjk{Figures \ref{fig:comparepsp}-\ref{fig:compareOMNI} show a comparison between the measured and GAMERA/ADAPT and GAMERA/AFT model in-situ data at Parker, SolO and Earth. For each time series, we overplot the GAMERA/ADAPT and GAMERA/AFT model predictions as black and red curves, respectively, as well as the associated SCS magnetic field data for times when Parker is below $R_\odot$. Time periods including interplanetary CMEs (ICMEs; these directly hit the spacecraft) in the HELIO4CAST catalog \citep{Moestl23} are marked in the shaded red regions.} \par 
\kjk{For each satellite, the plot of $v_R$ also includes a prediction from WSA/ADAPT or WSA/AFT \citep{Arge00}, which applies a constant radial velocity between $21.5 R_\odot$ and 1 AU to each plasma parcel originating at $21.5 R_\odot$ (where the velocity was obtained via \autoref{WSA}). At increments of $1/10$ AU, stream interaction regions modify this velocity via
\beg{SIRwsa}
v_R(R) = \sqrt{2}\Big[\big(v_R(R)\big)^{-2}+\big(v_R(R+dR)\big)^{-2}\Big]^{-1/2},
\done 
such that plasma parcels moving faster than the parcels ahead of them are slowed down. The polarity of the magnetic field assigned to each parcel is kept constant throughout space, allowing the sign of the magnetic field to be obtained at each satellite from WSA/ADAPT or WSA/AFT. We did not find significant differences in the results if this model used \autoref{SIRwsa} as opposed to assuming a constant velocity between $R_{in}$ and 1 AU. }

\begin{figure*}[ht]
\includegraphics[trim={0cm 0cm 0cm 0cm},clip,scale=0.65]{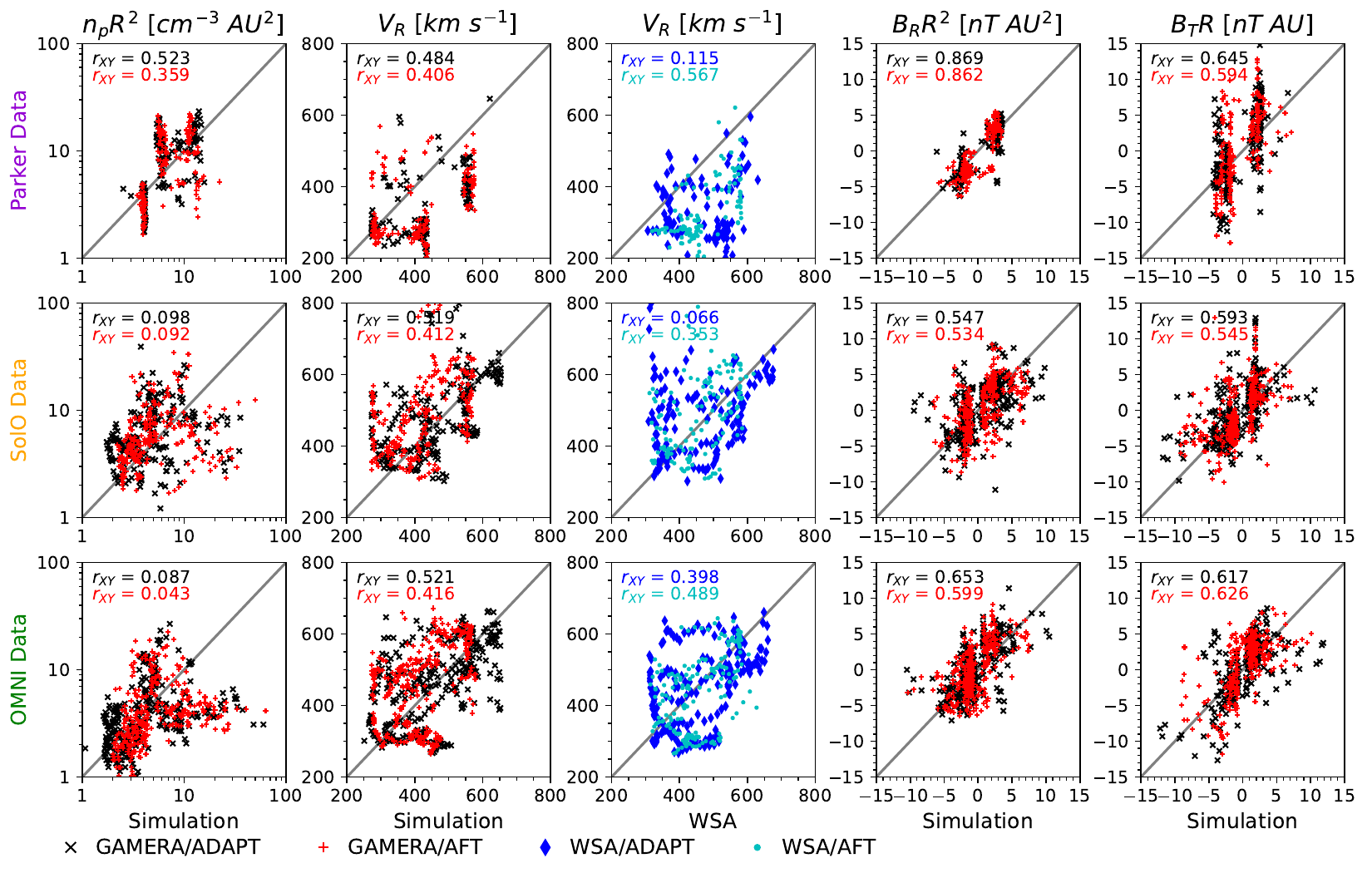}
\caption{Scatter plots comparing the observed and simulated $n_p$, $v_R$, $B_R$ and $B_T$ values for Parker (top row), SolO (middle row), and OMNI (bottom row). Black "x's" show the comparison for GAMERA/ADAPT results and red crosses show the GAMERA/AFT results. The third column shows the WSA/ADAPT and WSA/AFT model results for the radial velocity. In each plot, $r_{XY}$ denotes the sample Pearson correlation coefficient for each set of points.}\label{fig:obs_corr}
\end{figure*}

The shaded region seen near 01 Jun is due to the trajectory of Parker being inside $R_{in}$, so \kjk{we use the magnetic field obtained from the SCS layer of the model in this region.} For Parker, the general behavior of the radial and transverse magnetic field data is reproduced reasonably well with both GAMERA/ADAPT and GAMERA/AFT models. \kjk{Furthermore, although Parker is inside the inner boundary of our simulation near 01 Jun, the SCS layer of the models reproduce the first HCS crossing, as does the inferred polarity from the WSA models. Both MHD models capture the second and third HCS crossings, with GAMERA/ADAPT capturing the timing of the HCS crossing slightly better than GAMERA/AFT, while the inferred WSA/ADAPT and WSA/AFT polarities lags.} $B_N$ is always close to 0 in the GAMERA model since it is set to near $0$ by the initial conditions and the only new $B_N$ results from the $v_N$ component of \autoref{induction} generated by Parker being slightly out of the \kjka{equatorial plane}. \kjk{\citep{Mostafavi22} also find a small nonzero $v_N$ generated near stream interaction regions. The rest of the model's variables are qualitatively hard to compare against the in-situ measurements due to the number of data gaps. The model velocity structure has some regions of overlap with the in-situ data, but generally shows high speed solar wind throughout the time interval, even where the in-situ measurements show slow wind. A statistical analysis of the extant data (described below) does not show good agreement for density or radial velocity.} \par 
The same comparisons for SolO and OMNI data are shown in \autoref{fig:compareSolO} and \autoref{fig:compareOMNI}, respectively. \kjk{The MHD models seem to capture much of the qualitative behavior of the in-situ data. Qualitatively, the purely WSA predictions seem to perform worse than the MHD predictions at SolO, and comparably at Earth. At SolO, in particular, the MHD models seem to capture the crossings of the HCS a bit better than pure WSA, and reproduce the velocity better.} \par 

\kjk{We quantify the fit between our models and the in-situ data from each satellite by examining the Pearson coefficient between the simulated and observed plasma variables. In \autoref{fig:obs_corr} we show the observed and simulated $n_p$, $v_R$, $B_R$ and $B_T$ values for Parker (top row), SolO (middle row), and OMNI (bottom row). Results from the GAMERA/ADAPT run are plotted with black ``x's" while results from the GAMERA/AFT run are plotted with red crosses. In addition, the third column shows the radial velocity predicted using the WSA/ADAPT and WSA/AFT models (\autoref{SIRwsa}). These plots exclude the data during the interval when ICMEs were present, as our steady state model is not expected to reproduce the data during this time period. The results show that the MHD models perform best at matching Parker radial magnetic field data, with a Pearson $r_{XY}\sim0.86$. The radial magnetic field data is reasonably well reproduced at SolO ($r_{XY}\sim0.54$) and Earth ($r_{XY}\sim0.63$), but the density ($r_{XY}\sim0.1$) is not. The radial velocity at all three locations is somewhat poorly reproduced by the MHD models ($r_{XY}\sim0.4-0.5$), but WSA/ADAPT and WSA/AFT perform comparably or worse ($r_{XY}\sim0.1$) in all three cases. We also performed a phase offset study to see if shifting the model data in time by some arbitrary amount to the left or right (backward or forward in time) would be a better fit to the in-situ measurements, but the results indicated that the phase offset that maximized the Pearson coefficient was at most a few hours in all cases, and did not make a noticeable difference in the quality of the correlation.}\par 
\kjka{An additional feature of these quantitative comparisons is that the simulated radial velocity seems to be uniformly overpredicted at Parker but not at SolO or Earth. One possible explanation for this is that there is residual acceleration of the solar wind that happens past Parker \citep{Dakeyo22} that is not accounted for in WSA, which has fine-tuned its parameters for 1 AU.}

\subsection{Magnetic Connectivity}

\begin{figure*}[ht]
  \begin{minipage}{1.0\linewidth}
    \centerline{\includegraphics[clip,width=\linewidth]{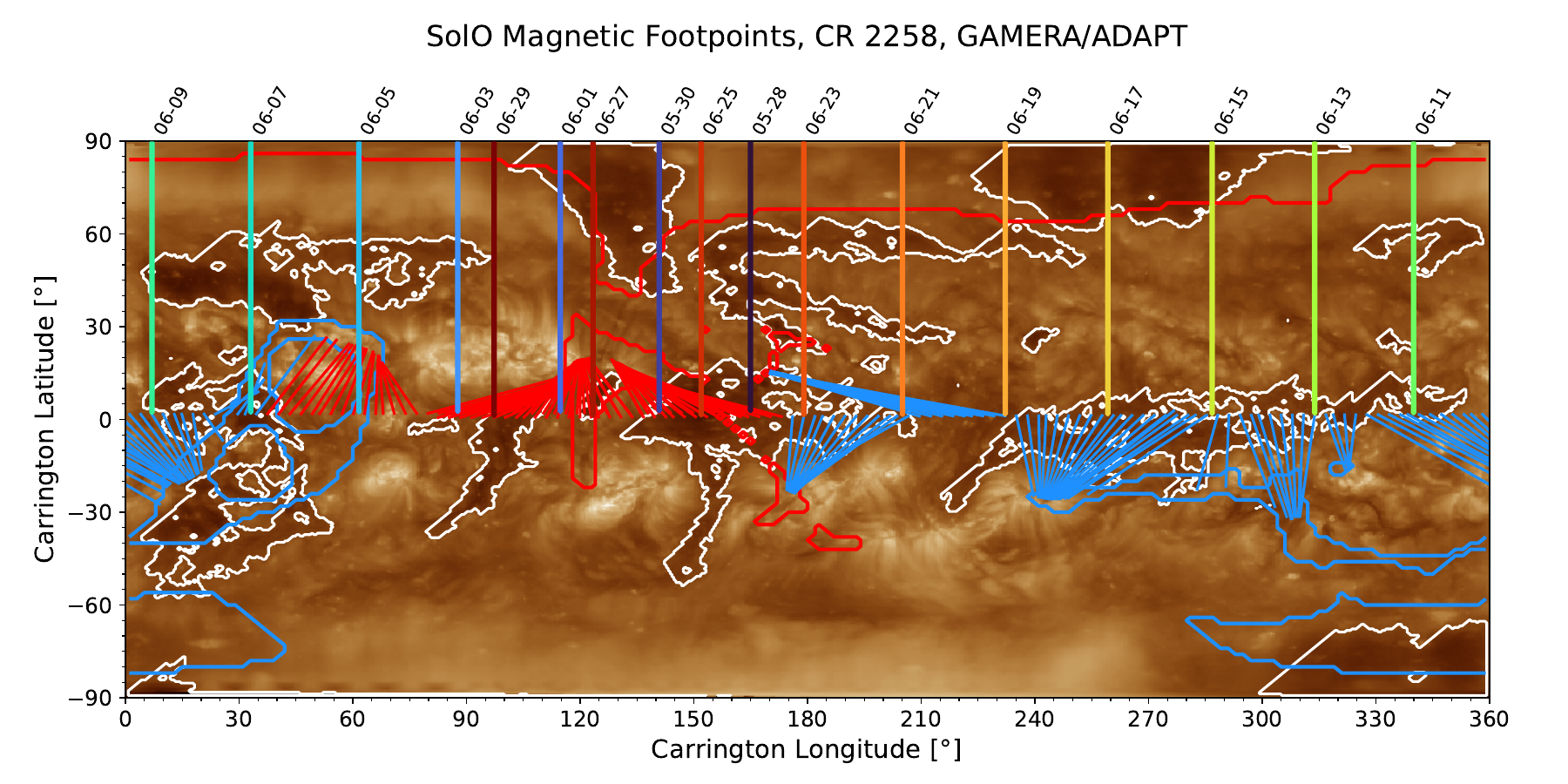}}
    \centerline{\includegraphics[clip,width=\linewidth]{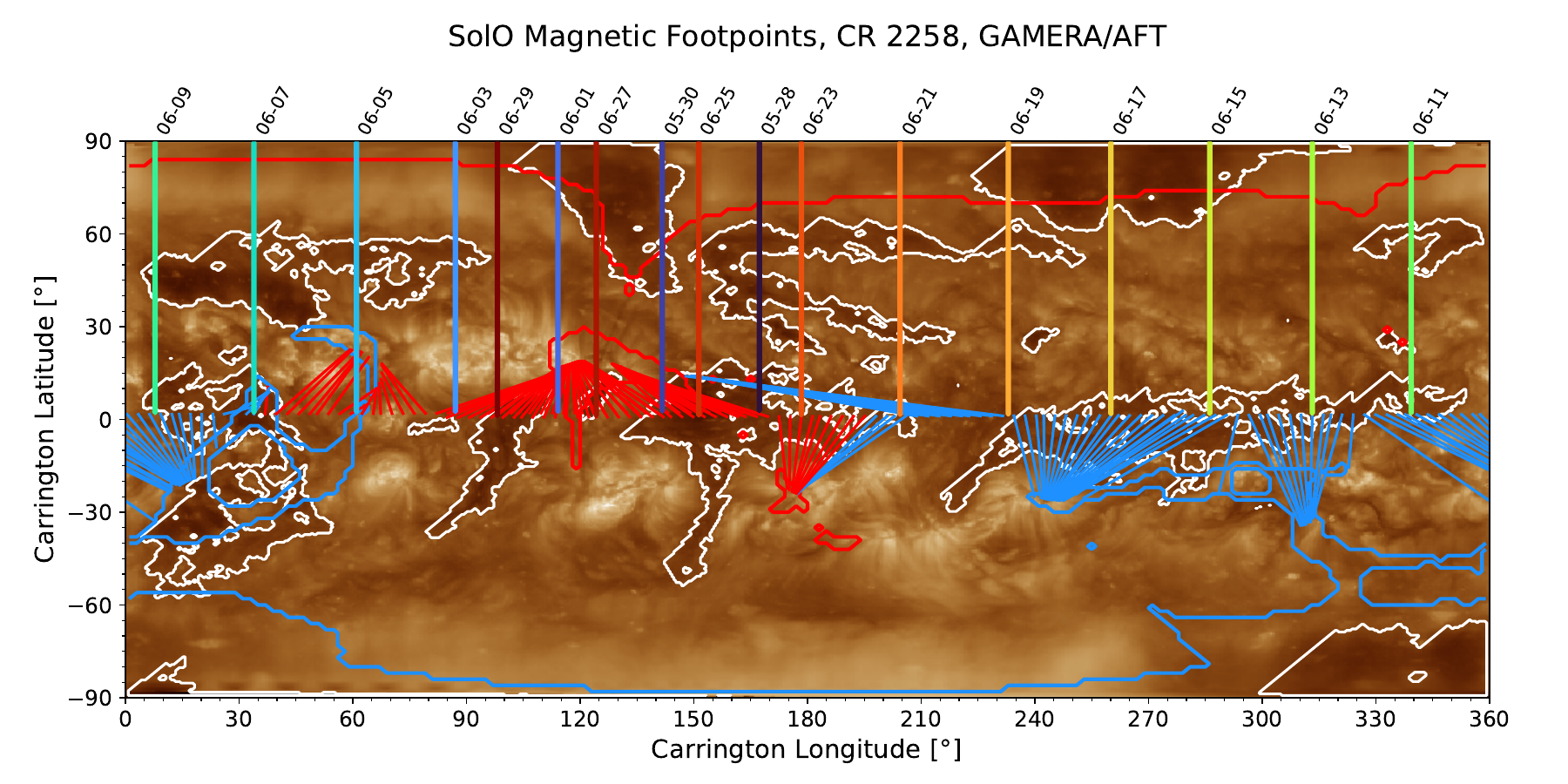}}
  \end{minipage}
\caption{The photospheric footpoints of the field lines connecting to \kjk{SolO} from both GAMERA models are overplotted on the associated synoptic AIA maps. Color shading of the thick vertical lines represents time, and the thin straight lines starting near $0^\circ$ latitude originate at the location of \kjk{SolO} and connect down to the photosphere. Although the field lines follow a curved trajectory, only the start and end points are shown, connected by a straight line, since the actual trajectory is hard to discern for so many field lines. The red/blue contours encircle open flux of positive/negative polarity on the photosphere based on the associated ADAPT-PFSS  and AFT-PFSS extrapolations. The measured in-situ sign of $B_R$ is coded on this plot in the color of the field lines. \kjk{White contours indicate coronal holes identified using the EZSEG algorithm.}}\label{fig:connectivity}
\end{figure*}

\begin{figure*}[htp]
  \begin{minipage}{1.0\linewidth}
    \centerline{\includegraphics[clip,width=\linewidth]{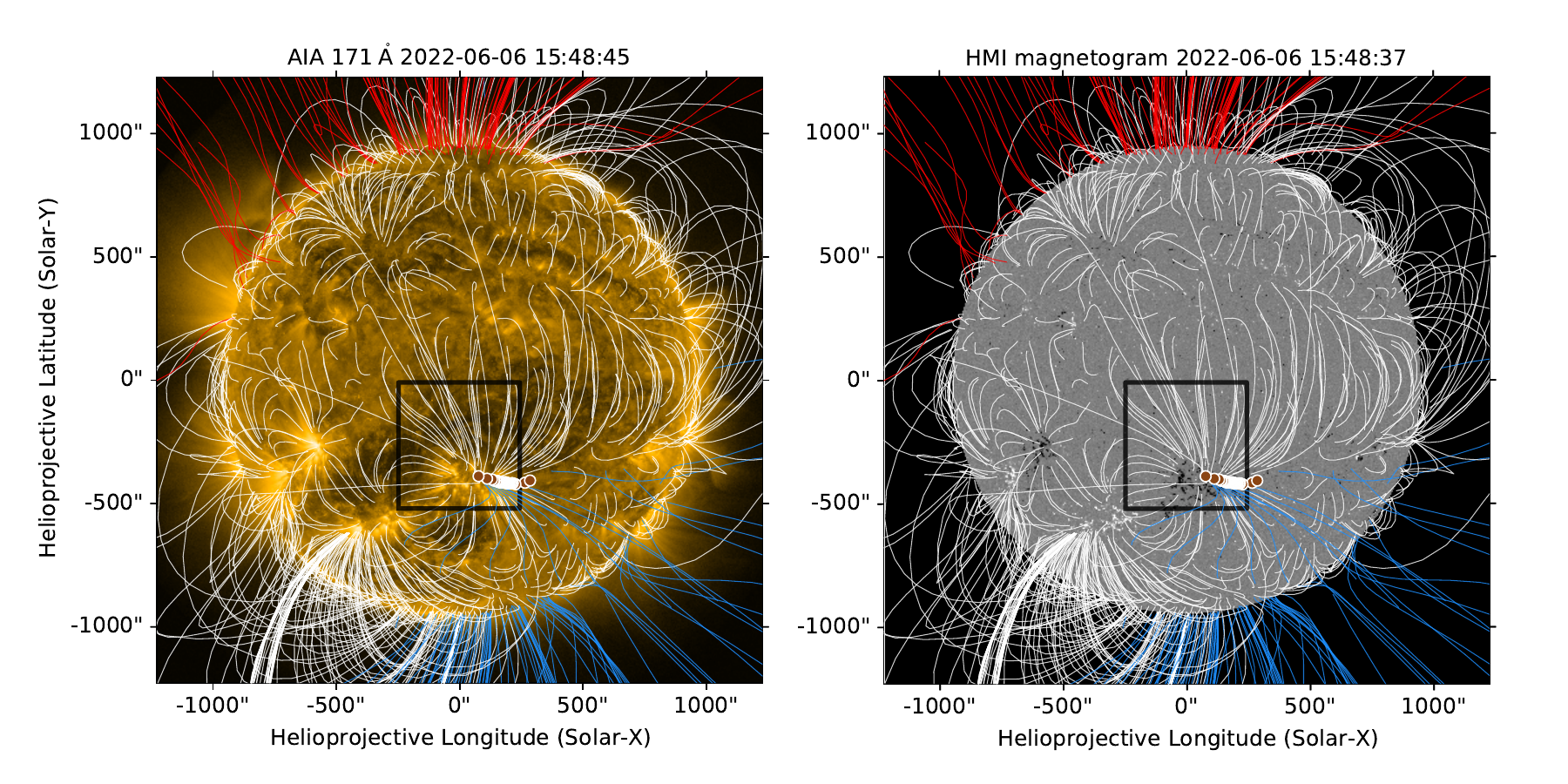}}
    \centerline{\includegraphics[clip,width=\linewidth]{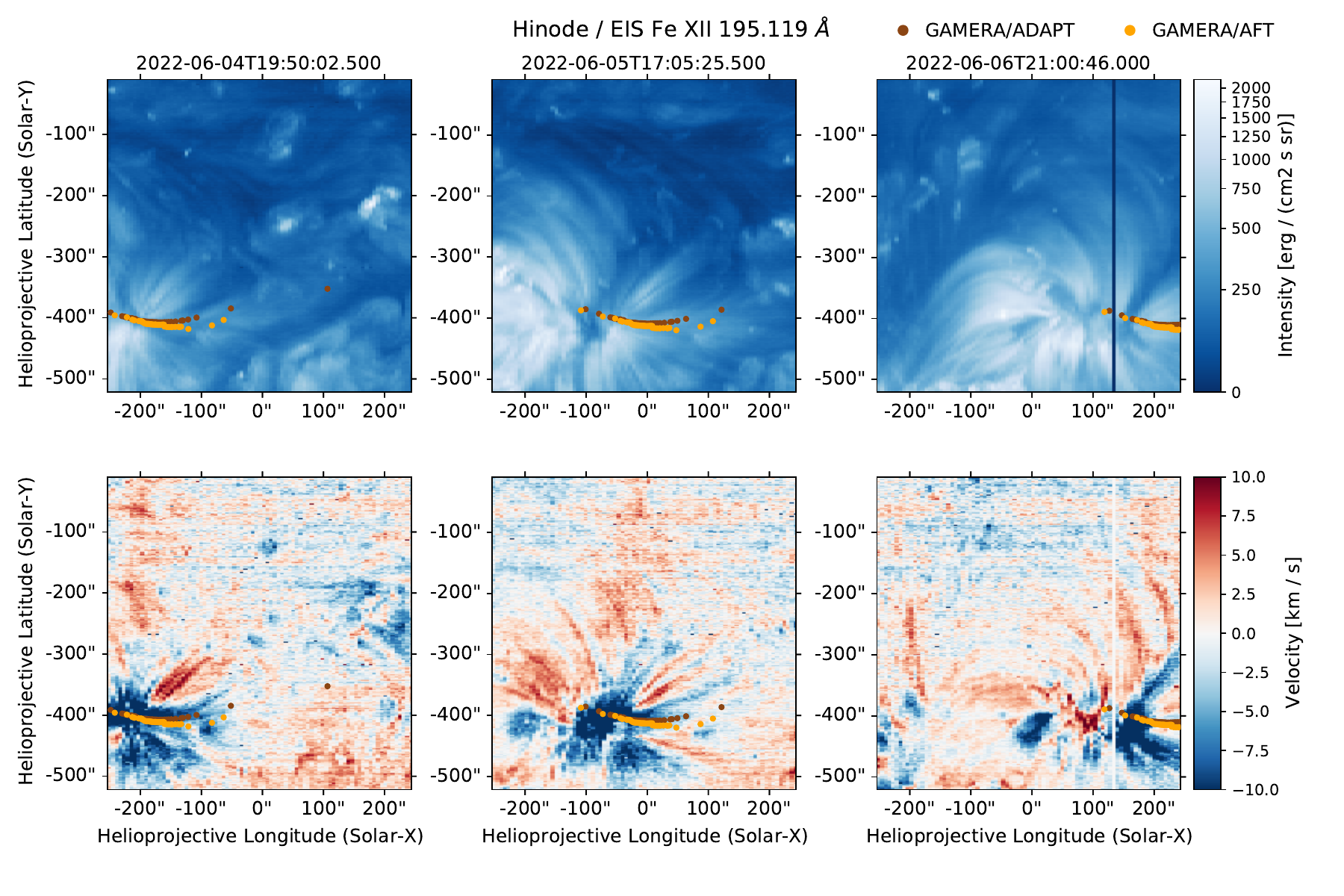}}
  \end{minipage}
\caption{\textit{Top panels:} PFSS field lines, calculated using a full-sun synoptic map, plotted on an AIA 171\,\AA\ image and HMI
  line-of-sight magnetogram. Closed field lines are shown in white. Open field lines are shown in
  red (positive) and blue (negative). The black box shows the EIS field of view. In the helioprojective coordinate system \citep{Thompson06}, the Cartesian variables $X$ and $Y$ correspond, respectively, to the longitude and latitude measured relative to the center of the observed solar disk.
  \textit{Middle panel:} Intensity and \textit{Bottom panel:} Doppler velocity measured by EIS in the 
  \ion{Fe}{12} 195.119\,\AA\ line at three times during this interval. The orange/brown dots shown are footpoints connected to SolO rotated to the time of the EIS raster from the ADAPT/AFT MHD models. In the bottom panel, Red/blue 
  indicates a downflow/upflow. \kjk{This convention, where velocities toward the observer are blue shifted with negative values, effectively reverses the direction of positive velocity with respect to the rest of the paper. We keep this standard convention in this Figure, and note that the negative velocity values here correspond to outflows away from the Sun.}}\label{fig:eis}
\end{figure*}

\kjk{The magnetic connectivity of the satellites to the photosphere can be tracked throughout the simulation in one of two ways. Either (1) ballistically \citep{Badman20}, where the $\phi$ position of the footpoint at $r=R_\odot$ is calculated at constant $\theta$ via
\beg{ballisticfieldline}
\phi(r=R_\odot) = \phi - \frac{180\Omega}{\pi v_R(r)}\Big(R_\odot - r\Big) ,
\done 
or (2) directly, by tracing magnetic field lines as described above, from the satellites to the inner boundary of GAMERA, whereupon the connectivity to the photosphere is determined by tracing the field lines through the SCS and PFSS solutions. Using these methods, we can obtain the modeled photospheric footpoint of the magnetic field line connected to each satellite at each time step.} \par
\kjk{We plot the locations of these footpoints calculated from method (2) for SolO in \autoref{fig:connectivity}. The footpoints are clearly seen to connect to the modeled open field areas (red and blue contours) in both the ADAPT and AFT maps, though these open field regions do not always coincide with the dark EUV emission areas (white contours) identified in the AIA 193 \AA  $\;$ images with the EZSEG algorithm \citep{Caplan16}, which uses an image segmentation technique combined with a simple region-growing algorithm. The mismatch between the bright regions and open field regions can sometimes result from bright coronal loops in the vicinity masking small coronal holes. The ballistic mapping footpoints are not quantitatively different from those obtained by direct tracing, in agreement with the finding that footpoint mapping is consistent across different combinations of coronal and heliospheric models \citep{Badman22}.} Occasionally the polarity of the traced field lines does not match the polarity of their source regions, as, for example, around \kjk{5 June and 23 June}. 


\kjk{During the time period 18--24 June 2022, both direct and ballistic connectivity models show that SolO was magnetically connected to an active region on the photosphere.}
We compare the locations of the photospheric
footpoints traced from \kjk{SolO} against Doppler maps obtained from spectroscopic
observations of the upper solar atmosphere, measured by calculating the Doppler shift along the line of sight in the Fe XII line measured with EIS. There have been many observations of active region outflows
with EIS that are suggestive of a connection to the solar wind
\citep[e.g.,][]{Sakao:2007,Del-Zanna:2008,Doschek:2008,Brooks:2011,Brooks:2015,Harra21}. As is illustrated
in Figure~\ref{fig:eis}, the \kjk{GAMERA/ADAPT and GAMERA/AFT field lines connecting SolO} to the solar surface suggest that the solar wind  \kjk{it measures during} the period of \kjk{14--19 June 2022} originated in an active region
outflow since there is a blue-shifted (negative velocity) plage region at the footpoints of magnetic field lines connected to SolO from both simulations. Although in this case there are open field lines associated with the outflow
region, this is not universally true, as some outflows are in regions where a PFSS model shows
closed field \citep[e.g.,][]{Culhane:2014}. It is not clear if, in these cases, the PFSS is a poor
representation of the magnetic topology or if the outflows are siphon flows on long, closed field
lines. 

\begin{figure*}[ht]
\includegraphics[trim={0cm 1cm 0cm 0cm},clip,scale=0.8]{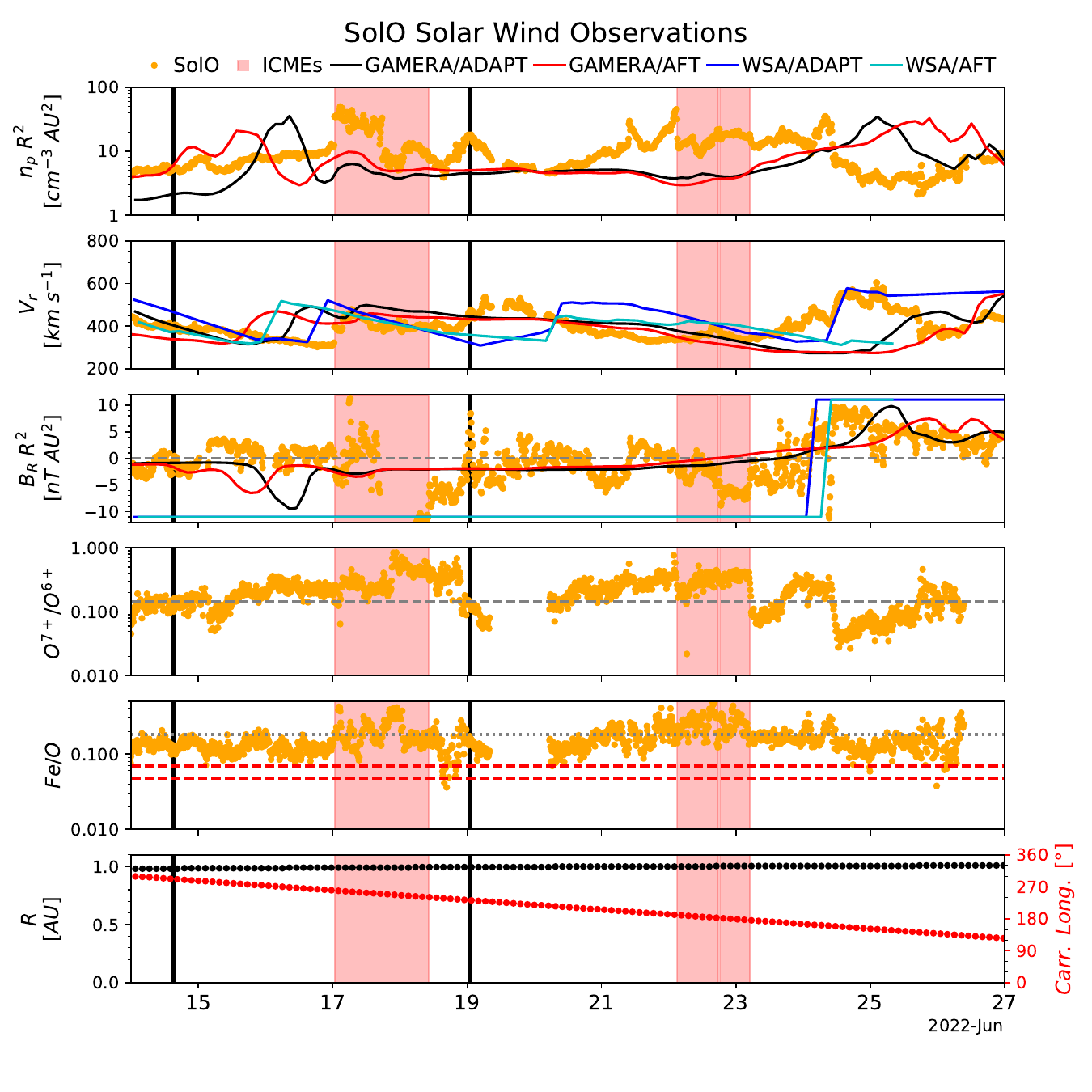}
\caption{Bulk solar wind observed by SolO / SWA-PAS ($n_p$ and $v_R$), MAG ($B_R$), and heavy ion ratios from SWA-HIS ($O^{7+}/O^{6+}$ and $Fe/O$). The black, vertical lines indicate the boundaries of the time range when SolO is thought to be connected to an active region. The dashed line in the $O^{7+}/O^{6+}$ panel indicates a reference value of 0.145, which roughly separates hotter, coronal source regions ($\geq 0.145$) from cooler, photospheric ones \citep[$< 0.145$;][]{Zhao09}. The dotted gray line in the $Fe/O$ panel is a reference coronal value of $Fe/O = 0.18$ from \cite{Feldman92} and the two dashed red lines correspond to photopsheric values of $Fe/O = 0.069$ and $0.047$ from \cite{Asplund09} and \cite{Grevesse98}, respectively.}\label{fig:zoomSolO}
\end{figure*}

\kjk{If the solar wind observed by SolO during 14--19 June 2022 originated in or near an active region, we would expect some signatures in the charge states and abundance ratios of heavy ions (e.g., He, C, O, Fe). In particular, we should expect to see higher $O^{7+}/O^{6+}$ ratios, indicative of a hotter source region, and abundance ratios closer to coronal values than photospheric \citep{Owocki83}. \autoref{fig:zoomSolO} shows a closer view of solar wind observed at SolO for 14--27 June. The top three panels show the bulk solar wind density and radial velocity measured by SolO's SWA-HIS, and radial magnetic field measured by the MAG instrument, along with the corresponding simulation results, while the fourth and fifth panels show the $O^{7+}/O^{6+}$ and $Fe/O$ heavy ion ratios measured by the SWA-PAS \citep{Louarn20,Owen20}. The dashed line at $0.145$ in the $O^{7+}/O^{6+}$ panel roughly separates coronal hole ($< 0.145$) and non coronal hole ($\geq 0.145$) type wind \citep{Zhao09}. The dotted grey line in the $Fe/O$ panel is a reference coronal value of $0.18$ from \cite{Feldman92} and the two dashed red lines are photopsheric values for $Fe/O$ of $0.069$ and $0.047$ from \cite{Asplund09} and \cite{Grevesse98}, respectively. Although several ICMEs passed SolO during this time period, there is, nevertheless, evidence of an increase in the heavy ion ratios of $O^{7+}/O^{6+}$ and $Fe/O$ over their reference (dashed line) values. While not definitive, this does provide evidence that SolO was, indeed, connected to an active region at this time.}

\section{Discussion}\label{sec:conclusions}
\kjk{In this paper, we presented the results of a comparison between heliospheric modeling using the ADAPT and AFT surface flux transport models. For both ADAPT and AFT maps for CR2258, we used a WSA velocity equation coronal model to initialize a GAMERA MHD heliospheric model. We compared the output of these models to in-situ data from Parker, SolO, and OMNI, and quantitatively compared the performance of GAMERA/ADAPT to GAMERA/AFT. We found that the models performed similarly in all cases, and that if one model matched (did not match) the in-situ data, the other model also matched (did not match). The models did best at matching the radial magnetic field at Parker, and did poorly at matching the radial velocity at all three satellites. The MHD simulations performed comparably to a purely WSA model at matching the radial velocity at Parker and Earth, but somewhat better at SolO. Finally, we performed a connectivity analysis to determine that SolO was magnetically connected to an active region observed by Hinode/EIS, and on-board instruments on SolO measured increased heavy ion ratios during this time, potentially indicating a hotter, AR, coronal, rather than cooler, coronal hole, photospheric, source region.}\par 
\kjk{Our work represents two novel additions to the literature. First, to our knowledge, this is the first coronal+heliospheric model to be performed using AFT. Previous work has used AFT to predict polar fields and the solar cycle \citep{Upton18} or tested ADAPT against other surface flux transport models \citep{Schrijver01} and found large differences when a new active region appeared in the assimilation window \citep{Barnes23}. However, this study assesses how well AFT performs as the input for coronal and heliospheric modeling relative to the industry standard of ADAPT, and finds that AFT performs just as well. Second, this is the first study, to our knowledge, that has attempted to reproduce in-situ SolO measurements using any coronal or heliospheric model. An intriguing result is the comparison between the WSA prediction of the velocity and the in-situ measurements at SolO. Given that SolO is nearly at 1 AU and very close to the equatorial plane during this time period, it could be expected that WSA should perform equally well at predicting the solar wind velocity at Earth and SolO. The poor performance of WSA at SolO is therefore somewhat surprising. One possibility is that WSA is correctly calibrated for 1 AU, but the far side magnetogram information is lacking, resulting in poor predictions $180^\circ$ from Earth, where SolO is situated.} \kjka{Interestingly, the simulated overestimate of the velocity seen at Parker, but not at SolO or Earth, indicates that the parameters of the WSA models, which have been fine tuned for 1 AU, are not appropriate for locations closer to the Sun. While this does seem to demonstrate the smaller \kjkaa{velocity} of the solar wind near the Sun \citep{Dakeyo22}, attempts to reset the parameters of WSA based on Parker data have not, thus far, yielded significantly improved results \citep{Samara22b}}.\par
\kjk{The overall results presented here are in line with previous heliospheric models -- both based purely on ballistic or the WSA velocity equation \citep{Arge00} and MHD approaches \citep{Kim20,Riley21}, or a mix thereof \citep{Badman22} -- that steady state models can reproduce in-situ properties at various positions in the heliosphere. Still, our work is the first to examine whether using the AFT flux transport model could improve heliospheric in-situ predictions over those made using ADAPT. Although AFT does not perform better than ADAPT, the fact that it performs comparably is evidence that the treatment of flux transport itself may not be the cause of discrepancies between model and in-situ measurements. Nevertheless, it seems likely that the source of the discrepancies between model predictions and in-situ measurements is in the coronal portion of these models. \kjka{Eight} suspect assumptions made in the coronal portion of our model include (1) The steady-state assumption (the Sun is dynamically evolving) and the lack of accurate far side information, (2) the PFSS assumption \citep[there are very likely currents between the photosphere and $R_{ss}$,][]{Schuck22}, (3) the SCS assumption (there is no reason to suppose that all currents above the photosphere are concentrated in a shell at $R_{ss}$ and in the HCS, (4) the WSA velocity assumption (which is an empirical relationship and does not account for mass or momentum conservation), (5) an empirical relationship between the velocity and the density at $R_{in}$ (which again does not account for mass or momentum conservation), \kjka{(6) the lack of a proper energy treatment, including heat flux and Alfv\'en wave damping, (7) the lack of time dependence of the boundary conditions, and (8) the lack of solar far-side information}. Coronal MHD models avoid \kjk{many of these} assumptions \citep[e.g.,][]{Riley12,Lionello14}, and assumption (3) (and potentially 2) seems to be vindicated (though not proven) by the generally good agreement of the model and in-situ magnetic field components, both in the heliospheric portion of the model and in the SCS layer. However, all of these assumptions, and especially the empirical relationships between the magnetic field, velocity, and density, are worth examining further.} \par 
\kjka{Another important takeaway from this work is that using SolO to model the solar wind enables associating in-situ compositional data with the source region on the Sun through the connectivity mapping. Although the results are not definitive, they do seem to be mostly connectivity model independent, indicating that SolO could reliably be used to understand active region and solar wind heating.}\par 
Our work does not endeavor to address long-standing problems in heliospheric physics such as the `open flux problem' \citep{Linker17,Arge23}, the origin of switchbacks \citep{Tenerani20, Drake21, Pecora22}, or the source of the slow solar wind \citep{Antiochos11, Abbo16,Higginson17}. Nevertheless, the ultimate goal of heliospheric modeling is the accurate and timely prediction of geoeffective events. 
Future work will attempt to compare different input boundary conditions to determine how optimally to reproduce in-situ data, and compare different models \citep[e.g.,][]{Riley21,Wu20} against each other. Another possibility is to change the parameters of the expansion factor in the WSA model \citep{Wu20,Samara22b} to improve the agreement with the in-situ data. Additionally, including far side information in the photospheric flux transport model \citep{Upton19} may be a way to improve the quality of in-situ predictions.

\acknowledgments
This work was sponsored by the Office of Naval Research 6.1 program (KJK, HPW, MGL, IUU), NASA's Living with a Star program (HPW, IUU, LAU, MGL), NASA's Parker Solar Probe/WISPR program (KJK, MGL), and NASA's Hinode program (MJW, YKK, HPW, IUU). EP acknowledges the support of NRL`s contract N00173-22-2-C601.
The authors are grateful to Karl Battams, Nicholas Crump, Cooper Downs, Brendan Gallagher, Carl Henney, Peter MacNeice, Slava Merkin, Pete Riley, Roger Scott, Kareem Sorathia, and Yi-Ming Wang for helpful discussions and technical advice. The authors are also grateful to the anonymous referee, whose careful reading of, and comments on, the initial manuscript helped highlight several issues.
HPW acknowledges support from ISSI for team 463, “Exploring The Solar Wind in Regions Closer Than Ever Observed Before.” We acknowledge the NASA Parker Solar Probe Mission and the SWEAP team led by J. Kasper for the use of data. The FIELDS experiment on the Parker Solar Probe spacecraft was designed and developed under NASA contract NNN06AA01C. Hinode is a Japanese mission developed and launched by ISAS/JAXA, collaborating with NAOJ as a domestic partner, NASA and UKSA as international partners. Scientific operation of the Hinode mission is conducted by the Hinode science team organized at ISAS/JAXA. This team mainly consists of scientists from institutes in the partner countries. Support for the post-launch operation is provided by JAXA and NAOJ (Japan), UKSA (U.K.), NASA, ESA, and NSC (Norway). 

\appendix \label{sec:appendix}
\section{Potential Magnetic Field Inside a Spherical Shell}\label{app:intro}
We want to solve for a potential magnetic field in a spherical shell, i.e., one that satisfies
\beg{curlB}
\nabla\times\vecB = 0.
\done 
The solution is
\beg{BgradPsi}
\vecB = -\nabla \Psi.
\done 
For a solenoidal magnetic field, $\Psi$ therefore satisfies
\beg{Laplace}
\nabla^2\Psi = 0
\done 
whose general solution is, in spherical coordinates \citep{Jackson98}:
\beg{solsph}
\Psi(r,\theta,\phi) = \sum_{l=0}^{\infty}\sum_{m=-l}^{l}\Big[A_{lm}r^l+B_{lm}r^{-(l+1)}\Big]Y_{lm}\oftp.
\done 
The $Y_{lm}\oftp$ are spherical harmonics satisfying the orthogonality condition
\beg{Yortho}
\int_0^{2\pi}d\phi\int_0^\pi{d\theta \;\sin\theta\; Y^*_{l'm'}\oftp Y_{lm}\oftp} = \delta_{l'l}\delta_{m'm}.
\done 
The components of the magnetic field $B=-\nabla \Psi$ are then:
\beg{gradpsir}
B_r = -\pd{\Psi}{r} = -\sum_{l=0}^{\infty}\sum_{m=-l}^{l}\Big[lA_{lm}r^{l-1}-(l+1)B_{lm}r^{-(l+2)}\Big]Y_{lm}\oftp,
\done 
\beg{gradpsit}
B_\theta = -\frac{1}{r}\pd{\Psi}{\theta} = -\sum_{l=0}^{\infty}\sum_{m=-l}^{l}\Big[A_{lm}r^{l-1}+B_{lm}r^{-(l+2)}\Big]\pd{Y_{lm}\oftp}{\theta},
\done 
\beg{gradpsip}
B_\phi = -\frac{1}{r\sin\theta}\pd{\Psi}{\phi} = -\frac{1}{\sin\theta}\sum_{l=0}^{\infty}\sum_{m=-l}^{l}\Big[A_{lm}r^{l-1}+B_{lm}r^{-(l+2)}\Big] im Y_{lm}\oftp.
\done

\section{The Potential Field Source Surface Model}\label{app:pfss}
The potential field source surface \citep[PFSS; ][]{Altschuler69} model solves for the magnetic field $\vecB=-\nabla \Psi$ inside a region given by $R_1\leq r \leq R_2$, where the boundary condition at $R_1$ is the Neumann condition
\beg{bc1pfss}
\pd{\Psi}{r}(r,\theta,\phi)\at_{R_1} = -B_r(R_1,\theta,\phi)\hat{\textbf{r}},
\done 
and the boundary condition at $R_2$ is the Dirichlet condition
\beg{bc2pfss}
\Psi(R_2,\theta,\phi) = 0.
\done 
\autoref{bc2pfss} implies, from \autoref{solsph}, that 
\beg{Blmpfss}
B_{lm} = -A_{lm}R_2^{2l+1}.
\done 
In the PFSS, $B_\theta$ and $B_\phi$ are not constrained by their measured values at $R_1$. Thus, the general solution for the radial magnetic field component in the region $R_1\leq r \leq R_2$ is:
\beg{Brpfss1}
\begin{split}
\pd{\Psi}{r} = -B_r(r,\theta,\phi) &= \sum_{l=0}^{\infty}\sum_{m=-l}^{l}\Big[lA_{lm}r^{l-1}+(l+1)A_{lm}R_2^{2l+1}r^{-(l+2)}\Big]Y_{lm}\oftp \\
& = \sum_{l=0}^{\infty}\sum_{m=-l}^{l}A_{lm}Y_{lm}\oftp\Big(lr^{l-1}+lR_2^{2l+1}r^{-(l+2)}+R_2^{2l+1}r^{-(l+2)}\Big).
\end{split}
\done 
To get the coefficients $A_{lm}$, the orthogonality condition in \autoref{Yortho} can be employed by multiplying both sides of \autoref{Brpfss1} by $\sin\theta Y_{l'm'}^*\oftp$, evaluating at $r=R_1$ using boundary condition \autoref{bc1pfss}, and integrating over $\theta$ and $\phi$:
\beg{calcApfss1}
\begin{split} 
\int_0^\pi d\phi \int_0^{2\pi}{d\theta\sin\theta}Y_{l'm'}^*\oftp B_r(R_1,\theta,\phi)  & = \sum_{l=0}^{\infty}\sum_{m=-l}^{l}\Big(lR_1^{l-1}+lR_2^{2l+1}R_1^{-(l+2)}+R_2^{2l+1}R_1^{-(l+2)}\Big) \\
&\times A_{lm}\int_0^\pi d\phi \int_0^{2\pi}{d\theta\sin\theta}Y^*_{l'm'}\oftp Y_{lm}\oftp,  \\
\end{split} 
\done 
so that
\beg{calcApfss2}
\int_0^\pi d\phi \int_0^{2\pi}{d\theta\sin\theta}Y_{l'm'}^*\oftp B_r(R_1,\theta,\phi) = \sum_{l=0}^{\infty}\sum_{m=-l}^{l}\Big(lR_1^{l-1}+lR_2^{2l+1}R_1^{-(l+2)}+R_2^{2l+1}R_1^{-(l+2)}\Big) A_{lm}\delta_{l'l}\delta_{m'm}.
\done 
The Kroenecker $\delta$'s force all contributions to the double sums to vanish except $l=l'$ and $m=m'$, whence we can write (dropping, for simplicity of notation, the primes on $l$ and $m$):
\beg{calcApfss3}
\int_0^\pi d\phi \int_0^{2\pi}{d\theta\sin\theta Y_{lm}^*\oftp B_r(R_1,\theta,\phi)}    = A_{lm}\Big(lR_1^{l-1}+lR_2^{2l+1}R_1^{-(l+2)}+R_2^{2l+1}R_1^{-(l+2)}\Big).
\done 
This expression can be solved for $A_{lm}$ by rearranging:
\beg{Almpfss}
A_{lm} = \frac{a_{lm}}{lR_1^{l-1}+lR_2^{2l+1}R_1^{-(l+2)}+R_2^{2l+1}R_1^{-(l+2)}}
\done 
where we have defined
\beg{littlealm}
a_{lm} \equiv  \int_0^\pi d\phi \int_0^{2\pi}{d\theta\sin\theta Y_{lm}^*\oftp B_r(R_1,\theta,\phi)}.
\done 
The radial magnetic field in the volume, \autoref{Brpfss1}, can then be written as:
\beg{Brpfss}
\begin{split}
B_{r,pfss}(r,\theta,\phi) &=  \sum_{l=0}^{\infty}\sum_{m=-l}^{l}a_{lm}Y_{lm}\oftp \frac{lr^{l-1}+lR_2^{2l+1}r^{-(l+2)}+R_2^{2l+1}r^{-(l+2)}}{lR_1^{l-1}+lR_2^{2l+1}R_1^{-(l+2)}+R_2^{2l+1}R_1^{-(l+2)}} \\
& = \sum_{l=0}^{\infty}\sum_{m=-l}^{l}a_{lm}Y_{lm}\oftp\Big(\frac{R_1}{r}\Big)^{(l+2)}\Big(\frac{l(r/R_2)^{2l+1}+l+1}{l(R_1/R_2)^{2l+1}+l+1}\Big) \\
& \equiv \sum_{l=0}^{\infty}\sum_{m=-l}^{l}a_{lm}c_l(r)Y_{lm}\oftp.
\end{split}
\done 
For completeness, the other two components of the magnetic field can be derived from \autoref{gradpsit}-\autoref{gradpsip} as:
\beg{Btpfss}
B_{\theta,pfss}(r,\theta,\phi)  = \sum_{l=0}^{\infty}\sum_{m=-l}^{l} a_{lm} d_l(r)\pd{Y_{lm}\oftp}{\theta},
\done 
\beg{Bppfss}
B_{\phi,pfss}(r,\theta,\phi) = \sum_{l=0}^{\infty}\sum_{m=-l}^{l} \;i\;m\; a_{lm}d_l(r)\frac{Y_{lm}\oftp}{\sin\theta}. 
\done 
Here we have defined \citep[cf.;][]{Wang92, Song23}
\beg{clr}
c_l(r) \equiv \Big(\frac{R_1}{r}\Big)^{(l+2)}\Big(\frac{l(r/R_2)^{2l+1}+l+1}{l(R_1/R_2)^{2l+1}+l+1}\Big),
\done 
and
\beg{dlr}
d_l(r) \equiv -\Big(\frac{R_1}{r}\Big)^{l+2}\Big(\frac{(r/R_2)^{2l+1}-1}{l(R_1/R_2)^{2l+1}+l+1}\Big).
\done 
At $r=R_2$, $d_l(r) = 0 \;\forall \; l$, making $B_\theta(R_2,\theta,\phi)=B_\phi(R_2,\theta,\phi)=0$.  

\section{The Schatten Current Sheet Model}\label{app:scs}
In the \citet{Schatten71} model, the magnetic field $B=-\nabla \Psi$ inside the region $R_2\leq r < \infty$, is also assumed to be potential, so \autoref{Laplace} is solved again. This time, the boundary conditions on $\Psi$ are the Neumann boundary condition at $R_2$\footnote{The fact that the PFSS model uses a Dirichlet condition at $R_2$, while the Schatten model uses a Neumann boundary condition at $R_2$ creates a discontinuity in the value of $\Psi$. The result of the implementation of the Schatten model is the formation of a large current sheet at $R_2$, so $B\ne -\nabla\Psi$ at $R_2$, and the magnetic field is not irrotational there. The magnetic field is, therefore, not singular at $R_2$.}:
\beg{bc1scs}
\pd{\Psi}{r}(r,\theta,\phi)\at_{R_2} = |B_{r,pfss}(R_2,\theta,\phi)|\hat{\textbf{r}},
\done 
and the Dirichlet condition
\beg{bc2scs}
\Psi(r\to\infty,\theta,\phi) = 0.
\done 
The absolute value in \autoref{bc1scs} is a key feature of the Schatten model. After the general solution is calculated, locations connected by the magnetic field to locations on $R_2$ having $B_r(R_2,\theta,\phi)<0$ return to their original polarity. \par 
The general expression for the three magnetic field components given in \autoref{gradpsir}-\autoref{gradpsip} requires that, to satisfy \autoref{bc2scs}, we must have  
\beg{Almschatten}
A_{lm} = \delta_{l0},
\done 
since otherwise the magnetic field will blow up as $r\to\infty$. Thus, \autoref{gradpsir}-\autoref{gradpsip} become: 
\beg{gradpsirsch}
B_r =  \sum_{l=0}^{\infty}\sum_{m=-l}^{l}\Big[(l+1)B_{lm}r^{-(l+2)}\Big]Y_{lm}\oftp,
\done 
\beg{gradpsitsch}
B_\theta = -\sum_{l=0}^{\infty}\sum_{m=-l}^{l}\Big[B_{lm}r^{-(l+2)}\Big]\pd{Y_{lm}\oftp}{\theta},
\done 
\beg{gradpsipsch}
\begin{split}
B_\phi = -\sum_{l=0}^{\infty}\sum_{m=-l}^{l}\Big[B_{lm}r^{-(l+2)}\Big] im Y_{lm}\oftp.
\end{split}
\done 
The coefficients $B_{lm}$ are determined by multiplying \autoref{gradpsirsch} by $\sin\theta Y_{l'm'}^*\oftp$, evaluating at $R_2$ using \autoref{bc1scs}, and integrating over $\theta$ and $\phi$:
\beg{calcBlmsch1}
\begin{split} 
\int_0^\pi d\phi &\int_0^{2\pi}{d\theta\sin\theta}Y_{l'm'}^*\oftp |B_{r,pfss}(R_2,\theta,\phi)|  =\\ 
& \sum_{l=0}^{\infty}\sum_{m=-l}^{l}\Big[(l+1)B_{lm}R_2^{-(l+2)}\Big]\int_0^\pi d\phi\int_0^{2\pi}{d\theta\sin\theta}Y_{l'm'}^*\oftp Y_{lm}\oftp. \\
\end{split}
\done 
The orthonormality of the spherical harmonics in \autoref{Yortho} yields:
\beg{calcBlmsch2}
\int_0^\pi d\phi \int_0^{2\pi}{d\theta\sin\theta}Y_{l'm'}^*\oftp |B_{r,pfss}(R_2,\theta,\phi)|  =\sum_{l=0}^{\infty}\sum_{m=-l}^{l}(l+1)B_{lm}R_2^{-(l+2)}\delta_{l'l}\delta_{m'm}.
\done 
The Kroenecker $\delta$'s force all contributions to the double sums to vanish except $l=l'$ and $m=m'$, and we can once again drop the primes on $l$ and $m$ to write:
\beg{calcBlmsch3}
\int_0^\pi d\phi \int_0^{2\pi}{d\theta\sin\theta}Y_{lm}^*\oftp |B_{r,pfss}(R_2,\theta,\phi)|  =(l+1)B_{lm}R_2^{-(l+2)}.
\done 
This results in the expression for $B_{lm}$
\beg{calcBlm}
B_{lm} = \frac{b_{lm}}{(l+1)R_2^{-(l+2)}},
\done 
where we have defined
\beg{littleblm}
b_{lm} \equiv \int_0^\pi d\phi \int_0^{2\pi}{d\theta\sin\theta}Y_{lm}^*\oftp |B_{r,pfss}(R_2,\theta,\phi)|.
\done 
The expression for the three magnetic field components outside $R_2$ then becomes:
\beg{brsch}
B_{r,sch} =  \sum_{l=0}^{\infty}\sum_{m=-l}^{l}g_l(r)b_{lm}Y_{lm}\oftp,
\done 
\beg{btsch}
B_{\theta,sch} = -\sum_{l=0}^{\infty}\sum_{m=-l}^{l}h_l(r)\;i\;m\;b_{lm}\pd{Y_{lm}\oftp}{\theta},
\done
\beg{bpsch}
B_{\phi,sch} = -\sum_{l=0}^{\infty}\sum_{m=-l}^{l}h_l(r)\;i\;m\;b_{lm}Y_{lm}\oftp,
\done
where
\beg{glr}
g_l(r) \equiv \Big(\frac{R_2}{r}\Big)^{(l+2)},
\done 
and
\beg{hlr}
h_l(r) \equiv \Big(\frac{R_2}{r}\Big)^{(l+2)}\frac{1}{l+1}.
\done 
Note also that the transformation $R_2 \to \infty$ and $R_1\to R_2$ makes $c_l(r) = g_l(r)$ and $d_l(r) = h_l(r)$. In this way, the Schatten model reduces to yet another implementation of the PFSS, with the outer boundary at infinity. \par 
At this stage, all fields rooted in negative $B_r$ at $R_2$ are reversed back to their original polarities. This requires tracing numerous field lines to obtain the sign of the field in the region $r>R_2$. This key step is the source of the heliospheric current sheet in the Schatten current sheet model, as the change in the direction of the magnetic field produces a large $\nabla\times\vecB$, creating a large current sheet separating positive and negative polarity regions.\par 
An important point about the Schatten model is that while $B_r$ is continuous across $R_2$, $B_\theta$ and $B_\phi$ are not, since setting $r=R_2$ in \autoref{btsch}-\autoref{bpsch} of the Schatten model does not make $B_\theta$ and $B_\phi$ vanish. This is incompatible with the imposed boundary condition at $R_2$ in the PFSS, where $B_\theta$ and $B_\phi$ are explicitly set to 0. There is, therefore, a nonzero $B_\theta$ and $B_\phi$ at $R_2$ in the Schatten model, while the PFSS model results in $B_\theta=B_\phi=0$ at $R_2$. Mathematically, this results from using a Dirichlet condition at $R_2$ when solving the Laplace equation in the region $R_1<r<R_2$, but a Neumman condition at $R_2$ when solving the same equation in the region $r>R_2$. Physically, this occurs because a potential field requires sources (currents) outside the domain of interest \citep{Schuck22}, which the PFSS places inside or at the photosphere and outside the source surface. The Schatten model puts all of the sources (currents) inside the source surface and at the heliospheric current sheet. These sets of currents are fundamentally inconsistent with the idea that the only current is in the heliospheric current sheet. In fact, there is another current sheet at $R_2$, so this setup cannot result in a continuous solution across the source surface for all magnetic field components. 
\par
\begin{figure*}
\includegraphics[trim={0cm 0cm 0cm 0cm},clip,scale=0.7]{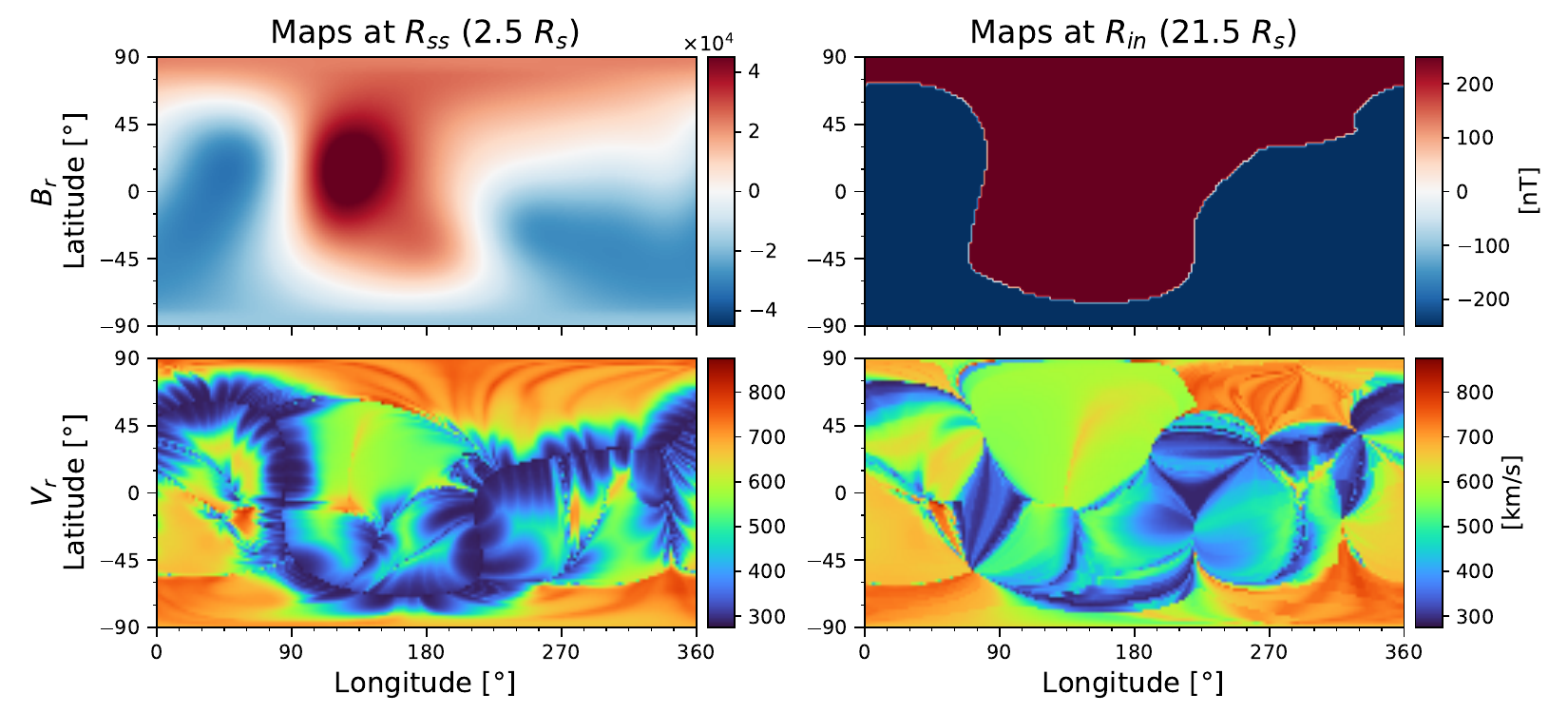}
\caption{Radial magnetic field (top) and radial velocity (bottom) calculated from the PFSS and WSA velocity equation models using ADAPT data, respectively, at the source surface (left) and as extrapolated to the GAMERA inner boundary (right). The right panels are the same as in \autoref{fig:inputmaps}, though without the open flux contours in the $B_r$ panel, and are shown for comparison.}\label{fig:schatten}
\end{figure*}

 
To deal with this issue, several authors \citep{Schatten69,Reiss19} perform a minimization procedure to minimize the least squared residual between the field at $R_2$ obtained by the SCS model (\autoref{brsch}-\autoref{bpsch}) and that determined from the PFSS. This process changes the coefficients $a_{lm}$ and $b_{lm}$ from those determined above in a way that minimizes the error in $B_r$, $B_\theta$ and $B_\phi$. This procedure is described in detail in \citet{Schatten69} and \citet{Reiss19}. In this work, we do not perform such a minimization, though it may be the subject of future improvements for this model. Instead, we take the radial magnetic field at the source surface from the PFSS model (shown in \autoref{fig:schatten}, top left) to calculate the Schatten model. Then we take the radial field at $r=21.5 R_\odot$ from the Schatten model as the inner boundary condition for GAMERA (shown in \autoref{fig:schatten}, top right), with $B_\theta=0$ and $B_\phi$ specified by \autoref{Bphi_init}. The value of $B_{\theta,sch}$ and $B_{\phi,sch}$ from \autoref{btsch} and \autoref{bpsch} are found to be about two orders of magnitude smaller than the value of $B_{r,sch}$ from \autoref{brsch}, so this approximation can be justified \emph{a posteriori}. The velocity at the source surface (computed on a regular grid from \autoref{WSA}) and at the GAMERA inner boundary is also shown in the bottom panels of \autoref{fig:schatten} for context. The velocity on the inner boundary is plotted by tracing field lines from a regular grid on the inner boundary down to the source surface, where the velocity has been computed from \autoref{WSA}, and taking the values of the velocity at the locations of the field line footpoints on the source surface.

\newpage 
\software{
Astropy \citep{Astropy18},
h5py \citep{h5py},
IPython \citep{IPython07},
Matplotlib \citep{Matplotlib07},
Numpy \citep{Numpy06},
Pandas \citep{pandas},
Scipy \citep{Scipy20},
SpiceyPy \citep{Annex20}.
}
\newpage 
\bibliographystyle{aasjournal}
\providecommand{\noopsort}[1]{}\providecommand{\singleletter}[1]{#1}%

\end{document}